\documentclass[12pt,a4paper]{article}

\usepackage{amsmath,amsfonts,latexsym,amssymb}
\usepackage{verbatim,amsthm,curves,graphics}
\usepackage{mathrsfs}
\usepackage{feynmp}

\def\ben{\begin{equation}}
\def\een{\end{equation}}
\def\bena{\begin{eqnarray}}
\def\eena{\end{eqnarray}}

\def\f(#1/#2){\frac{#1}{#2}} 
\def\Frac(#1/#2){\left(\frac{#1}{#2}\right)} 
\def\chris(#1-#2-#3){{\mit \Gamma}^{#1}{}_{{#2}{#3}} }
\def\tilchris(#1-#2-#3){\tilde{{\mit \Gamma}}^{#1}{}_{{#2}{#3}}}
\def\hatchris(#1-#2-#3){\hat{{\mit \Gamma}}^{#1}{}_{{#2}{#3}}}

\theoremstyle{definition}

\newtheorem{lem}{Lemma}
\newtheorem{prop}{Proposition}

\newcommand{\W}{{\mathcal W}}
\newcommand{\Z}{{\mathcal Z}}
\renewcommand{\S}{{\mathscr S}}
\newcommand{\inv}{{\rm inv}}
\newcommand{\B}{{\mathcal A}}
\newcommand{\V}{{\mathcal V}}
\newcommand{\s}{{\underline s}}
\newcommand{\supp}{{\rm supp}}
\newcommand{\X}{{\mathcal X}}
\newcommand{\C}{{\mathcal C}}

\newcommand{\mr}{{\mathbb R}}
\newcommand{\mc}{{\mathbb C}}
\newcommand{\WF}{{\rm WF}}
\newcommand{\D}{{\mathscr D}}

\newcommand{\tr}{{\rm Tr \,}}
\newcommand{\A}{{\mathcal A}}

\newcommand{\e}{{\rm e}}
\newcommand{\E}{{\mathcal E}}
\newcommand{\eps}{{\varepsilon}}
\newcommand{\myid}{{\bf 1}}
\renewcommand{\Phi}{{\mathcal O}}

\begin{document}

\title{Algebraic Approach to the $1/N$ Expansion in Quantum Field Theory} 
\author{Stefan Hollands\thanks{Electronic mail: \tt stefan@bert.uchicago.edu}\\
                    \it{Enrico Fermi Institute, Department of Physics,}\\
                    \it{University of Chicago, 5640 Ellis Ave.,}\\
                    \it{Chicago IL 60637, USA} \\
        }
\date{\today}

\maketitle 

\begin{abstract}
The $1/N$ expansion in quantum field theory is formulated within an algebraic framework. 
For a scalar field taking values in the $N$ by $N$ hermitian matrices, 
we rigorously construct the gauge invariant 
interacting quantum field operators in the sense of power series in 
$1/N$ and the `t Hooft coupling parameter as members of an abstract *-algebra. 
The key advantages of our algebraic formulation over the usual formulation of 
the $1/N$ expansion in terms of Green's functions are (i) that it 
is completely local so that infra-red divergencies in massless theories are avoided on 
the algebraic level and (ii) that it admits a generalization to quantum field theories 
on globally hypberbolic Lorentzian curved spacetimes.
We expect that our constructions are also applicable in models possessing 
local gauge invariance such as Yang-Mills theories.

The $1/N$ expansion of the renormalization
group flow is constructed on the algebraic level via a family of *-isomorphisms between
the algebras of interacting field observables corresponding to different scales. We 
also consider $k$-parameter deformations of the interacting field algebras that arise
from reducing the symmetry group of the model to a diagonal subgroup with $k$
factors. These parameters smoothly interpolate between situations of different symmetry.
\end{abstract}

\section{Introduction}

A common strategy to gain (at least approximate) information about physical models 
is to expand quantities of interest in terms of the parameters of the model. For example 
in perturbation theory, one expands in terms of the coupling parameter(s) of the theory. In quantum 
theories, it is sometimes fruitful to expand in terms of Planck's constant, $\hbar$. The key 
point in all cases is that the theory one is expanding about---often a linear theory in the first 
example, and the classical limit in the second example---is under better control, and that 
there exist, in many cases, systematic and constructive schemes to calculate the deviations order by order. 
Another such expansion that has by now become standard in quantum field theory is the 
expansion in $1/N$, where $N$ describes the number of components of the field(s) in the model. As in the 
previous two examples, the theory that one expands about, i.e., the large $N$ limit, is often somewhat 
simpler than the theory at finite $N$, and can sometimes even be solved exactly. 
(Just as an example, it can happen~\cite{p} that the large $N$ limit of a non-renormalizable 
theory is renormalizable, and the $1/N$-corrections remain renormalizable.)

The $1/N$ expansion in quantum field theory was first introduced by `t Hooft~\cite{th} in the context 
of non-abelian gauge theories. He observed that, if one explicitly keeps track of all factors of $N$
in the perturbative expansion of a connected Green's function of gauge invariant interacting fields, then 
the series can be organized as a power series in $1/N$, provided that the coupling parameter of the theory
is also chosen to depend on $N$ in a suitable way. Moreover, he showed that the Feynman diagrams 
associated with terms at a given order in $1/N$ can naturally be related to Riemann surfaces with 
a number of handles equal to that order. Thus, in the $1/N$ expansion of this model, 
the leading contribution corresponds to planar diagrams, the subleading contribution to diagrams 
with a toroidal topology, etc.

The usual schemes for calculating the Green's functions in perturbation theory 
implicitly assume that the interacting fields approach suitable ``in''-fields in the asymptotic past, 
which one assumes can be identified with the fields in the underlying free field theory that one is expanding about. 
The existence of such ``in''-fields is closely related with the possibility to interpret the theory 
in terms of particles, and with the existence of an $S$-matrix. However, none of these usually exist in 
massless theories. Thus, the formulation of the $1/N$ expansion in terms of Green's functions
is potentially problematical in massless theories.
These problems come into even sharper focus if one considers theories on non-static 
(globally hypberbolic) Lorentzian spacetimes. Here there is not, in general, available 
even a preferred vacuum state based on which to calculate the Green's functions. Moreover, 
one would certainly not expect the fields to approach free ``in''-fields in the asymptotic past 
for example in spacetimes that do not have suitable static regions in the asymptotic future and past
(such as our very own universe). 

A strategy to avoid these difficulties has recently been developed in~\cite{bf,df}. The new idea 
in those references is to construct directly
the interacting field {\em operators} as members of some *-algebra of observables, rather 
than trying to construct the Green's functions. 
The key advantage of this approach is that, as it turns out, the interacting fields operators 
can always be defined in a completely satisfactory way,
without any reference to an imagined (in general non-existent) ``in''-field, 
or ``in''-states in the asymptotic past. 
Consequently, infra-red problems do not arise on the level of the interacting 
field observables and their associated algebras\footnote{Such divergences will arise, if 
one tries to construct (non-existent) quantum states in the theory corresponding to ``free
incoming particles''.}.  
Not surprisingly, these ideas have also been a key ingredient in the construction of 
interacting quantum field theories in curved spacetimes~\cite{hw2,hw3,hw1}. 

The purpose of this article is to show that it is also possible to formulate the $1/N$ expansion directly
in terms of the interacting fields and their associated algebras of observables to which these fields belong, thereby 
achieving a complete disentaglement from the infra-red behavior of the theory. 
We will consider in this article explicitly only the theory of 
a scalar field in the ${\bf N} \otimes \bar {\bf N}$ 
representation of $U(N)$ on Minkowski space. However, since our algebraic construction is
done without making use of any of the particlular features of Minkowski space, 
it can be generalized to arbitrary Lorentzian curved spacetimes 
by the methods of~\cite{bf,hw2,hw3}. Also, we expect that our methods are applicable to models 
with {\em local} gauge symmetry such as Yang-Mills theories, although the algebraic construction 
of the interacting fields in such 
theories is more complex due to the presence of unphysical degrees of freedom, and 
still a subject of investigation, see e.g.~\cite{df2, hs, hks}. 

\medskip

We now summarize the contents of this paper. In section~2 we review, in a pedagogical way, the 
construction of the field observables and the corresponding algebra associated with a single, free hermitan
scalar field $\phi$. This algebra is sufficiently large to contain 
the Wick powers of $\phi$ and their time ordered products, which are required later in the construction of 
the corresponding interacting quantum field theory. A description 
of the properties of these objects and their construction is therefore included. 

In section~3, we generalize
these constructions to a free scalar field in the ${\bf N} \otimes \bar {\bf N}$ representation of $U(N)$, and 
we show how to construct the $1/N$ expansion of this theory on the level of field observables and their associated algebras. 

In section~4, 
we proceed to interacting quantum field theories. Based on the algebraic construction of the underlying free quantum
field theory in section~3, we construct the interacting quantum fields as formal power series in $1/N$ and the 
`t Hooft coupling as members of a suitable algebra $\B_V$, where $V$ is a gauge invariant 
interaction. The contributions to these quantities arising at order
$H$ in the $1/N$ expansion correspond precisely to Feynman diagrams whose 
toplogy is that of a Riemann surface with $H$ 
handles. We point out that the algebras $\B_V$ also incorporate 
an expansion in $\hbar$ (``loop-expansion'')~\cite{df}, 
as well as by construction an expansion the coupling parameters appearing in $V$. Therefore,  
the construction of $\B_V$ in fact incorporates all three expansions mentioned at the beginning. 

In section~5 we review, first for a single scalar field, 
the formulation of the renormalization group in the algebraic framework~\cite{hw1} that we are working in. 
We then show that this construction can be generalized to 
an interacting field in the ${\bf N} \otimes \bar {\bf N}$ representation of $U(N)$
with gauge invariant interaction, defined in the sense of a power series in $1/N$. The renormalization group map, 
is therefore also defined as a power series in $1/N$. 

In section~6, we vary the constructions of 
sections~3 and~4 by considering interactions that are invariant only under some subgroup of $U(N)$. We show 
that, for a diagonal subgroup with $k$ factors, the algebraic construction of the $1/N$ expansion can 
still be carried through, but now leads to deformed algebars of interacting field observables which are 
labeled by $k$ real deformation parameters associated with the relative size of the subgroups. These parameters smoothly interpolate 
between situations of different symmetry.
The contributions to an interacting field at order $H$ in $1/N$ are now associated 
with Riemann surfaces that are ``colored'' by $k$ ``spins'', where each coloring is weighted according 
to the values of the deformation parameters. 

\section{Algebraic construction of a single scalar quantum field}

The perturbative construction of an interacting quantum field 
theory is based on the construction on the corresponding free quantum field
theory, and we shall therefore begin by considering free fields. 
In this section, we will review
how to define an algebra of observables associated 
with a single free hermitian Klein-Gordon field of mass $m$, described by the 
the classical action\footnote{Our signature convention is $-+++\dots$.} 
\ben
\label{1}
S = \int (\partial_\mu \phi \partial^\mu \phi + m^2 \phi^2) \, d^dx, 
\een
which is large enough in order to 
contain the Wick powers and the time ordered products of the field $\phi$. Our review is 
essentially self-contained and follows the ideas developed in~\cite{df,hw1,hw2,hw3,bf}, which the reader 
may look up for details. 

For pedagogical purposes, we begin by defining first a ``minimal algebra'' of observables associated with 
the action~\eqref{1}. Consider the free *-algebra over
the complex numbers generated by a unit $\myid$ and formal expressions $\phi(f)$ and 
$\phi(h)^*$, where $f$ and $h$ run through the space of compactly supported 
smooth testfunctions on $\mr^d$. The minimal algebra is obtained 
by factoring this free algebra by the following relations. 
\begin{enumerate}
\item
(linearity) $\phi(af + bh) = a\phi(f) + b\phi(h)$ 
for all $a, b \in \mc$ and testfunctions $f, h$. 
\item
(field equation) $\phi((\partial^\mu \partial_\mu - m^2)f) = 0$ for 
all testfunctions $f$. 
\item
(hermiticity) $\phi(f)^* = \phi(\bar f)$. 
\item
(commutation relations)
$[\phi(f), \phi(h)] = i\Delta(f, h) \cdot \myid$, where $\Delta(f, h)$ is the advanced
minus retarded propagator for the Klein-Gordon equation, smeared 
with the testfunctions $f$ and $h$.
\end{enumerate}

We formally think of the expressions $\phi(f)$ as the ``smeared'' 
quantum fields, i.e., the integral of the formal\footnote{Since $\Delta$ is 
a distribution, relation 4)
implies that the field $\phi$ necessarily has a distributional 
character. Therefore, the field only makes good mathematical sense
after smearing with a test function.} pointlike quantum 
field against the test function $f$, 
\ben
\label{2}
\phi(f) = \int_{\mr^d} \phi(x) f(x) \, d^d x. 
\een 
The linearity of the expression $\phi(f)$ in $f$ corresponds
to the linearity of the integral. Relation 2) is 
the field equation for $\phi(x)$ in the sense of distributions, i.e., 
it formally corresponds to the Klein-Gordon equation for $\phi(x)$ via a partial integration. 
Relation 3) says that the field $\phi$ is hermitian and relation 4) 
implements the usual commutation relations of the 
free hermitian scalar field on $d$-dimensional Minkowski spacetime. 

The minimal algebra is too small for our purposes. It does not, for example,  
contain observables corresponding to Wick powers of the field $\phi$ at 
the same spacetime point, nor their time ordered products. These are, however, required 
if one wants to construct the interacting quantum field theory perturbatively 
around the free theory. We now construct an enlarged algebra, $\W$, 
which contains elements corresponding to these observables.  
For this purpose, it is useful to first present the minimal algebra in terms of a 
new set of generators, defined by $W_0 = \myid$, and
\ben
\label{3}
W_a(f_1 \otimes \cdots \otimes f_a)
= (-i)^a \frac{\partial^a}{\partial \lambda_1 \cdots \partial \lambda_a}
\e^{i\phi(F)} \e^{\frac{1}{2}\Delta_+ (F, F)} \bigg|_{\lambda_i = 0}, 
\quad F = \sum^a \lambda_i f_i, 
\een
where $\Delta_+$ is any distribution in 2 spacetime variables which 
is a solution to the Klein-Gordon equation in each variable, and which has  
the property that its antisymmetric part is equal to $(i/2) \Delta$. In particular, 
we could choose $\Delta_+ = (i/2) \Delta$ at this stage, but it is important to leave this 
choice open for later. It follows from the definition that $W_1(f) = \phi(f)$, that
the quantities $W_a$ are symmetric under exchange 
of the testfunctions, and that 
\ben
\label{4}
W_a(f_1 \otimes \cdots \otimes f_a)^* = W_a(\bar f_1 \otimes  \cdots \otimes f_a), \quad
W_a(f_1 \otimes \cdots  (\partial^\mu \partial_\mu - m^2)f_i \otimes \cdots f_a) = 0.
\een
Using the 
algebraic relations 1)--4), one can express the product 
of two such quantities again as a linear combination of such quantities, 
\begin{multline}
\label{5}
W_a(f_1 \otimes \cdots \otimes f_a) \cdot W_b(h_1 \otimes \cdots \otimes h_b) \\
= \sum_{\mathcal P} 
\prod_{(k, l) \in {\mathcal P}} \Delta_+(f_k, h_l) \cdot W_c(\otimes_{i \notin {\mathcal P}_1}
f_i \otimes \otimes_{j \notin {\mathcal P}_2} h_j), 
\end{multline} 
where the following notation has been used to organize 
the sum on the right side: We consider sets $\mathcal P$ 
of pairs $(i, j) \in \{1, \dots, a\} \times \{1, \dots, b\}$; 
we say that $i \notin {\mathcal P}_1$ if there is no $j$ such that
$(i, j) \in \mathcal P$, and we say  
$j \notin {\mathcal P}_2$ if there is no $i$ such that
$(i, j) \in \mathcal P$. The number $c$ is related to $a$ and $b$ 
by $a + b - c = 2|{\mathcal P}|$, where $|{\mathcal P}|$ is the 
number of pairs in $\mathcal P$.  

It is easy to see that relations~\eqref{4} and~\eqref{5} form an equivalent 
presentation of the minimal algebra, i.e., we could equivalently
{\em define} the minimal algebra to be the  
abstract algebra generated by the elements $W_a(\otimes_i f_i)$, 
subject to the relations~\eqref{4} and~\eqref{5}, instead of defining it
as the algebra generated by $\phi(f)$ subject to the relations
1)--4) above (this is true no matter what the particular choice 
of $\Delta_+$ is). Thus, all we have done so far is to rewrite
the minimal algebra in terms of different gerenators. 

To obtain the desired extension, $\W$, of the minimal algebra, we
now choose a distribution
$\Delta_+$ which is not only a bisolution to the Klein-Gordon 
equation with the property that its antisymmetric 
part is equal to $(i/2)\Delta$, but which has the additional 
property that that it is of positive frequency type in 
the first variable and of negative frequency type in the 
second variable. This condition is formalized by demanding 
that the wave front set\footnote{It 
can be shown that the wave front set of a 
distribution is actually invariantly defined as a subset of 
the cotangent space of the manifold on which the distribution is 
defined. The set~\eqref{6} should 
therefore be intrinsically thought of as a subset of $T^*(\mr^d \times \mr^d)$.} $\WF(\Delta_+)$, 
(for the definition of the wave front set of a distribution, see~\cite{h})
has the following, so-called ``Hadamard'', property
\begin{multline}
\label{6}
\WF(\Delta_+) \subset \{(x_1, x_2; p_1, p_2) \in (\mr^{d} \times \mr^d) \times (\mr^{d} \times \mr^d 
\setminus (0,0)) \mid \\
p_1 = -p_2, (x_1-x_2)^2 = 0, p_1 \in \bar V^+\} \equiv \C_+, 
\end{multline}
where $\bar V^\pm$ denote the closure of the future resp. past lightcone in $\mr^d$. 
The key point is now that the relations~\eqref{4} and \eqref{5} make sense not only 
for test {\em functions} of the form 
$f_1 \otimes \cdots \otimes f_a$, 
but even much more generally for any 
test {\em distribution}, $t$, in the space $\E'_a$ of compactly 
supported distributions
in $a$ spacetime arguments which have the property that 
their wave front set does not contain any element of the form 
$(x_1, \dots, x_n; p_1, \dots ,p_n)$ such that all 
$p_i$ are either in the closure of the forward lightcone, 
or the closure of the past light cone,  
\begin{multline}
\label{7}
\E'_a = \{t \in \D'(\times^a \mr^d) \mid 
\text{$t$ comp. supp., $\WF(t)$ has no element} \\
\text{in common with 
$(\times^a \mr^d) \times (\times^a \bar V^+)$
or $(\times^a \mr^d) \times (\times^a \bar V^-)$}\}.  
\end{multline}
Indeed, these conditions on the wave front set on the $t$, together
with the wave front set properties~\eqref{6} of $\Delta_+$ can be 
shown to guarantee that the potentially ill defined products of 
distributions occurring in a product 
$W_a(t) \cdot W_b(s)$ are in fact well-defined and are such that the 
resulting terms are each of the form $W_c(u)$, with $u$ again an element in
the space $\E'_{c}$.\footnote{
We note that the wave front set of $(i/2)\Delta$ 
is {\em not} of Hadamard type, and $\Delta_+ = (i/2) \Delta$ is not 
a possible choice in the construction of $\W$.}

We take $\W$ to be the algebra generated by 
symbols of the form $W_a(t)$, $t \in \E'_a$, subject to the relations~\eqref{4}
and~\eqref{5}, with $\otimes_i f_i$ in those relations replaced by 
distributions in the spaces $\E'_a$. Our definition of the 
generators $W_a(t)$ depends on the particular choice of $\Delta_+$. However, 
as an abstract algebra, $\W$ is independent of this choice~\cite{hw2}. To see this, choose any other 
bidistribution $\Delta_+'$ with the same wave front set property as $\Delta_+$, 
and let $\W'$ be the corresponding algebra with generators $W'_a(t)$ 
defined as in eq.~\eqref{3}. Then $\W$ and $\W'$ are isomorphic. 
The isomorphism is given in terms of the generators 
\ben
\label{8}
W_a(t) \to \sum_{2n \le a} \frac{a!}{(2n)!(a-2n)!} W_{a-2n}'(\langle F^{\otimes n}, t\rangle), 
\een
where $F = \Delta_+ - \Delta'_+$, 
and where $\langle F^{\otimes n}, t\rangle$ is the compactly supported distribution
on $\times^{a-2n} \mr^d$ defined by
\ben
\label{9}
\langle F^{\otimes n}, t\rangle(y_1, \dots, y_{a-2n}) = \int
t(x_1, \dots, x_{2n}, y_1, \dots, y_{a-2n}) \prod_i F(x_i, x_{i+1}) \, \prod d^d x_i. 
\een 
(That this distribution is in the class $\E'_{a-2n}$ follows from 
the wave front set properties of $\Delta_+, \Delta_+'$, which, together
with the wave equation imply that $F$ is smooth.) 

This completes our construction of the algebra of quantum observables for a single 
Klein-Gordon field associated with the action~\eqref{1}. Quantum {\em states} in 
the algebraic framework are by definition linear functionals $\omega: \W \to \mc$ which are positive in 
the sense that $\omega(A^*A) \ge 0$ for all $A \in \W$, and which are normalized so 
that $\omega(\myid) = 1$. This algebraic notion of a quantum state encompasses the 
usual Hilbert-space notion of state, in the sense that any vector or density matrix 
in a Hilbert-space on which the elements of $\W$ are represented as linear operators
defines an algebraic state in the above sense via taking expectation values. Conversely, given an 
algebraic state $\omega$, the GNS-construction yields a representation $\pi$ on a Hilbert space $\mathcal H$ 
containing a vector $|\Omega\rangle$ such that $\omega(A) = \langle \Omega | \pi(A) | \Omega \rangle$. 
Note, however, that it is not true that any state
on $\W$ arises in this way from a {\em single}, given Hilbert-space representation (this is closely
related to the fact that $\W$ has (many) inequivalent representations). The algebra $\W$ 
can be equipped with a unique topology that makes the product and *-operation continuous~\cite{hw2}, and
the notion of a {\em continuous} state on $\W$ can thereby be defined.  
The continuous states $\omega$ on $\W$ can be characterized entirely in terms of the
$n$-point distributions $\omega(\phi(x_1) \cdots \phi(x_n))$ 
of the field $\phi$. Moreover, it can be shown~\cite{hr}, that the continous states 
are precisely those for which the 2-point distribution has wave front set eq.~\eqref{6}, 
and for which the so-called ``connected''  $n$-point distributions are smooth for $n \neq 2$.

The invariance of the action~\eqref{1} under the Poincare-group is reflected in a 
corresponding invariance of $\W$, in the sense that $\W$ admits an automorphic action of the Poincare-group:  
For any element $\{\Lambda, a\}$ of the Poincare group consisting of a proper, orthochronous
Lorentz transformation $\Lambda$ and a translation vector $a \in \mr^d$, there is an 
automorphism $\alpha_{\{\Lambda, a\}}$
on $\W$ satisfying the composition law $\alpha_{\{\Lambda, a\}} \circ \alpha_{\{\Lambda', a'\}}
= \alpha_{\{\Lambda, a\} \cdot \{\Lambda', a'\}}$. The action of this automorphism 
is most easily described if we choose a $\Delta_+$ which is invariant under the Poincare-group. 
(Since $\W$ is independent of the choice of $\Delta_+$, we may do so if we like.) An admissible\footnote{That the wave front set of the Wightman
function is equal to~\eqref{6} is proved e.g. in \cite{rs}.} choice for $\Delta_+$ with the 
above properties is the Wightman function of the free field, 
\ben
\label{10}
\Delta_+(x, y) = w^{(m)}(x-y) \equiv \frac{1}{(2\pi)^{d-1}}
\int_{p^0 \ge 0} \delta(p^2 - m^2) e^{ip(x-y)} \, d^d p.  
\een
With this choice, the action of $\alpha_{\{\Lambda, a\}}$ is simply given 
by\footnote{That $t \circ {\{\Lambda, a\}}$ is again an element of $\E'_a$ is 
a consequence of the covariant transformation law $\WF(f^* t) = f^* \WF(t)$ of the 
wave front set~\cite{h}, where $f$ can be any diffeomorphism, together with the fact that the 
future/past lightcones are preserved under the action of the proper, orthochronous Poincare group.
On the other hand, a Lorentz transformation reversing the time orientation does not 
preserve the spaces $\E'_a$ and consequently does not give rise to an automorphism of $\W$.} 
\ben
\label{11}
\alpha_{\{\Lambda, a\}}(W_a(t)) = W_a(t \circ {\{\Lambda, a\}}).  
\een
Furthermore, with this choice for $\Delta_+$, 
the generators $W_a(t)$ correspond to the usually 
considered normal ordered products of fields, 
\ben
\label{12}
W_a(f_1 \otimes \cdots \otimes f_a) = \,: \prod^a_{i=1} \phi(f_i) :,  
\een
and the product formula~\eqref{5} simply corresponds to 
``Wick's theorem'' for multiplying to Wick-polynomials.

Since $W_1(f) = \phi(f)$, the enlarged algebra $\W$ contains
the minimal algebra generated by the free field $\phi(f)$ as a subalgebra. In fact, $\W$ also contains
Wick powers of the field at the same spacetime point as well as their time-ordered products, which 
are not in the minimal algebra. These objects can be characterized axiomatically 
(not uniquely, as we shall see) by a number of properties 
that we will list now. In order to state these properties of in 
a convenient way, let us introduce the vector space $\V$ whose basis 
elements are labelled by formal products of the field $\phi$ and 
its derivatives, 
\ben
\label{13}
\V = \text{span} \{ \Phi = \prod \partial_{\mu_1} \cdots \partial_{\mu_k} \phi \},  
\een
so that each element of $\V$ is given by  a formal linear combination
$\sum g_j \Phi_j, g_j \in \mc$. We refer to the elements of $\V$ as 
``formal'' field expressions, because no relations such as the field equation are 
assumed to hold at this stage. Consider, furthermore, the space 
$\D(\mr^d; \V)$ of smooth functions of compact support whose values 
are elements in the vector space $\V$. Thus, any element $F \in \D(\mr^d; \V)$
can be written in the form $F(x) = \sum f_i(x) \Phi_i$, where $\Phi_i$ 
are basis elements in $\V$, and where $f_i$ are complex valued 
smooth functions on $\mr^d$ of compact support.

We view the Wick powers as linear maps
\ben
\label{14}
\D(\mr^d; \V) \to \W, \quad f\Phi \to \Phi(f)
\een
and the $n$-fold time ordered products as multi linear maps 
\ben
\label{15}
T: \times^n \D(\mr^d; \V) \to \W, \quad
(f_1 \Phi_1, \dots, f_n \Phi_n) \to T(\prod f_i\Phi_i). 
\een
Time ordered products with only one factor are required to 
be given by the corresponding Wick power, 
\ben
T(f\Phi) \equiv \Phi(f). 
\een
The further properties 
required from the time ordered products (including the Wick powers as a 
special case) are the following\footnote{Actually, one ought to impose additional renormalization conditions specifically for 
time ordered products containing derivatives of the fields, see e.g.~\cite{df2} and \cite{hw4}, beyond the 
requirements (t1)--(t8) below. Such 
conditions are important
e.g. in order to show that the field equations or conservation equations hold for the interacting fields, but
they do not play a role in the present paper. We have therefore omitted them here to keep things as simple
as possible.}:
\begin{enumerate}
\item[(t1)] (symmetry) The time ordered products are symmetric under exchange of the arguments.
\item[(t2)] (causal factorization)
If $I$ is a subset of $\{1, \dots, n\}$, and if the supports of $\{f_i\}_{i \in I}$  
are in the causal future of the supports of $\{f_j\}_{j \in I^c}$ ($I^c$ denotes
the complement of $I$), then we ask that 
\ben
\label{16}
T\left(\prod_i f_i\Phi_i \right)= T\left(\prod_{i \in I} f_i\Phi_i \right) T \left(\prod_{j \in I^c} f_j \Phi_j
\right).  
\een 
\item[(t3)] (commutator)
\begin{multline}
\label{17}
\left[T\left(\prod_{i=1}^n f_i \Phi_i \right), \phi(h)\right] \\
= i\sum_k T\left(f_1\Phi_1 \cdots \sum_{\mu_1 \dots \mu_l}(h \partial_{\mu_1} \dots \partial_{\mu_l} \Delta * f_k)
\frac{\partial \Phi_k}{\partial(\partial_{\mu_1} \dots \partial_{\mu_l} \phi)}
 \cdots f_n \Phi_n \right),  
\end{multline}
where we have set $(\Delta*f)(x) = \int \Delta(x-y) f(y) \, d^d y$.
\item[(t4)] (covariance)
Let $\{\Lambda, a\}$ be a Poincare transformation. Then
\ben
\label{18}
\alpha_{\{\Lambda, a\}} \left( T(\prod_i f_i \Phi_i ) \right) =
T\left(\prod_i \psi^*_{\{\Lambda, a\}} (f_i \Phi_i ) \right),   
\een
where $\psi^*_{\{\Lambda, a\}}$ 
denotes the pull-back of an element in $\D(\mr^d; \V)$ by the linear transformation 
$x \to \Lambda x + a$.
\item[(t5)] (scaling) The time ordered products have the following ``almost homogeneous''
scaling behavior under simultaneous rescalings of the intertial coordinates and the mass, $m$. 
Let $\lambda > 0$, and set $f^\lambda(x) = \lambda^{-n} f(\lambda x)$ 
for any function or distribution on $\mr^n$. For a given prescription $T^{(m)}$ 
for the value $m$ of the mass (valued in the algebra $\W^{(m)}$ associated with 
this value of the mass), consider the new prescription $T^{(m) \prime}$ defined by 
\ben
\label{tlambda}
T^{(m) \prime} \left( \prod_i f_i \Phi_i \right) \equiv \lambda^{-\sum d_i} 
\sigma_\lambda \left[ T^{(\lambda m)} \left( \prod_i f_i^\lambda \Phi^{}_i \right) \right], 
\een
where $\sigma_\lambda: \W^{(\lambda m)} \to \W^{(m)}$ is the canonical isomorphism\footnote{
The canonical isomorphism is defined by $\sigma_\lambda: \W^{(\lambda m)} \owns W_a(t) \to 
\lambda^{-a} \cdot W_a(t^\lambda) \in \W^{(m)}$.}, and where $d_i$ is the ``engineering dimension''\footnote{
In scalar field theory in 
$d$ spacetime dimensions, the mass dimension of a field $\Phi$ is defined as 
the number of derivatives plus $(d-2)/2$ times the number of factors of $\phi$ plus 
twice the number of factors of $m^2$.
}
of the field $\Phi_i$. 
(Note that $T^{(\lambda m)}$ is valued in $\W^{(\lambda m)}$.)
Then we demand that $T^{(m) \prime}$ depends at most logarithmically on $\lambda$ in the sense that\footnote{
The difference $T^{(m)} - T^{(m) \prime}$ describes the failiure of $T_m$ to scale {\em exactly} homogeneously.
For the time ordered products with only one factor (i.e., the Wick powers), it can 
be shown that this difference vanishes, i.e., the Wick powers scale exactly homogeneously in 
Minkowski space. 
For the time ordered products with more than one factor, the logarithms cannot in general 
be avoided.}
\ben
\label{tlog}
T^{(m) \prime}= T^{(m)} + \text{polynomial expressions in $\ln \lambda$}. 
\een 
\item[(t6)] (microlocal spectrum condition)
Let $\omega$ be a continuous state on $\W$. Then the distributions $\omega_T: (f_1, \dots, f_n)
\to \omega(T(\prod f_i \Phi_i))$ are demanded to have wave front set
\begin{equation}
\WF(\omega_T) \subset \C_T, 
\end{equation}
where the set $\C_T \subset (\times^n \mr^d) \times (\times^n \mr^d \setminus \{0\})$ 
is described as follows (we 
use the graphological notation introduced in \cite{bfk,bf}):  
Let $\Gamma(p)$ be a ``decorated embedded Feynman graph''
in $\mr^d$. By this we mean an embedded Feynman graph $\mr^d$ whose 
vertices are points $x_1, \dots, x_n$ with valence specified by 
the fields $\Phi_i$ occurring in the time ordered product under consideration, 
and whose edges, $e$, are oriented null-lines [i.e., 
$(x_i - x_j)^2 = 0$ if $x_i$ and $x_j$ are connected by an edge]. Each such null 
line is equipped with a momentum vector $p_e$ parallel to that line.  
If $e$ is an edge in $\Gamma(p)$ connecting the points $x_i$ and $x_j$ 
with $i < j$, then $s(e) = i$ is its source 
and $t(e) = j$ its target. It is required that
$p_e$ is future/past directed if $x_{s(e)}$ is not in the 
past/future of  $x_{t(e)}$.
With this notation, we define
\begin{eqnarray}
\label{gamtdef}
\C_T &=& 
\Big\{(x_1, \dots, x_n; k_1, \dots, k_n) \mid 
\exists \,\, \text{decorated Feynman graph $\Gamma(p)$ with vertices} \nonumber\\
&& \text{$x_1, \dots, x_n$ such that
$k_i = \sum_{e: s(e) = i} p_e - \sum_{e: t(e) = i} p_e 
\quad \forall i$} \Big\}. 
\end{eqnarray}
\item[(t7)] (unitarity)
We have $T^* = \bar T$, where $\bar T$ is the ``anti-time-ordered
product´´, defined as
\begin{equation}
\label{unitary}
\bar T(f_1 \Phi_{1} \dots f_n\Phi_{n}) = 
\sum_{I_1 \sqcup \dots \sqcup I_j = \{1, \dots, n\}}
(-1)^{n + j} T( 
\prod_{i \in I_1} \bar f_i \Phi_i) \dots
T(\prod_{i \in I_j} \bar f_i \Phi_i), 
\end{equation}
where the sum runs over all partitions of the set $\{1, \dots, n\}$ into
disjoint subsets $I_1, \dots, I_j$.

\item[(t8)] (smooth dependence upon $m$)
The time ordered products depend smoothly upon the mass parameter $m$ in the following sense.
Let $\omega^{(m)}$ be a 1-parameter family of states on $\W^{(m)}$. We say that 
$\omega^{(m)}$ depends smoothly 
upon $m$ if (1) the 2-point function $\omega^{(m)}_2(x,y)$ when viewed 
as a distribution jointly in $m, x, y$ has wave front set
\ben
\WF(\omega^{(m)}_2) \subset \{ (x_1, x_2, m; p_1, p_2, \rho) \mid (p_1, p_2, \rho) \neq 0,(x_1, x_2; p_1, p_2) \in \C_+ \}, 
\een
where the set $\C_+$ was defined above in eq.~\eqref{6}, and if (2) 
the truncated $n$-point functions $\omega^{(m) {\rm conn}}_n$ are 
smooth jointly in $m, x_1, \dots, x_n$. We say the prescription $T^{(m)}$ is smooth 
in $m$ if $\omega_T^{(m)}(f_1, \dots, f_n) = \omega^{(m)} (T^{(m)}(\prod f_i \Phi_i))$ (viewed as a distribution 
jointly in $m$ and its spacetime arguments) has wave front set
\begin{multline}
\WF(\omega^{(m)}_T) \subset\Big\{(x_1, \dots, x_n, m; k_1, \dots, k_n, \rho) \mid   \\
(k_1, \dots, k_n, \rho) \neq 0, (x_1, \dots, x_n; k_1, \dots, k_n) \in \C_T \}
\Big\} 
\end{multline}
for such a smooth family of states, where the set $\C_T$ was defined above in eq.~\eqref{gamtdef}.

\end{enumerate}

It is relatively straightforward to demonstrate the existence of 
a prescription for defining the Wick powers as elements of $\W$ 
satisfying the above properties. For example, for the fields 
$\phi^a \in \V$, $a = 1, 2, \dots$, the corresponding algebra elements
$\phi^a(f) \in \W$ satisfying the above properties may be defined as follows. 
Let $H^{(m)}(x, y)$ be any family of bidistributions satisfying the 
wave equation in both entries and the wave front set condition~\eqref{6}, 
whose antisymmetric part is equal 
to $(i/2) \Delta(x,y)$, and which has a smooth dependence upon $m$ in the 
sense of (t8). Define
\ben
\label{wickpowers}
\phi^a(x) = \frac{\delta^n}{i^n \delta f(x)^n} e^{i \phi(f) + \frac{1}{2} H^{(m)}(f, f)} \Bigg|_{f=0}.
\een
Then $\phi^a(f) \in \W$ satisfies (t1)--(t8). This definition of $\phi^a(f)$ 
can be restated equivalently as follows: We may use the bidistribution
$\Delta_+ = H^{(m)}$ in the definition of the generators $W_a$ (see eq.~\eqref{3})
and the algebra product~\eqref{5} of $\W$, since we have already argued that $\W$ is independent
of the particular choice of $\Delta_+$. Consider the distribution $t$ given by  
\ben
\label{19}
t(x_1, \dots, x_a) = 
f(x_1) \delta(x_1 - x_2) \cdots \delta(x_{a-1} - x_a),
\een 
where $\delta$ is the ordinary delta-distribution in $\mr^d$. Then one 
can show that $t$ is in the class of distributions $\E'_a$, and definition~\eqref{wickpowers}
(in smeared form) is equivalent to setting
\ben
\label{20}
\phi^a (f) = W_a(t) \in \W, 
\een
where it is understood that $W_a$ is defined in terms of  
$\Delta_+ = H^{(m)}$. Wick powers containing derivatives 
are defined in a similiar way via suitable derivatives of delta distributions.

The usual ``normal ordering'' prescription for Wick powers would correspond to 
setting $H^{(m)}$ equal to the Wightman 2-point function $w^{(m)}$ given above 
in eq.~\eqref{10}. However, this is actually not an admissible choice in our framework
since, by inspection $w^{(m)}$ (and hence the vacuum state) does {\em not} depend smoothly upon $m$ in the sense of 
(t8). In fact, the Wightman 2-point function $w^{(m)}(x-y)$ explicitly contains a term of the form 
$J[m^2 (x-y)^2] \log m^2$ with a logarithmic dependence upon the mass $m$, 
where $J$ is a smooth (in fact, analytic) function that can 
be expressed in terms of Bessel functions. For this reason, the usual normal ordering
prescription violates our condition (t8) that the Wick powers 
have a smooth dependence upon $m$. An admissible choice for $H^{(m)}$ is 
e.g. $w^{(m)}$ without this logarithmic term, 
\ben
H^{(m)}(x, y) = w^{(m)}(x-y) - J[m^2 (x-y)^2] \log m^2.
\een
Since normal ordering is not admissible in 
our framework, it follows that no prescription for Wick powers satisfying (t1)--(t8) can 
have the property that it has a vanishing expectation value in the vacuum state for {\em all}
values of $m \in \mr$, because this property precisely distinguishes normal ordering. However, 
we can always adjust our prescription within the freedom left over by (t1)--(t8) in such 
a way that all Wick powers have a vanishing expectation value in the vacuum state for 
an arbitrary, but {\em fixed} value of $m$. It is therefore clear that, in practice, our prescription 
is just as viable as the usual normal ordering prescription, since $m$ can take on only one
value. On the other hand, our prescription would lead to different predictions in a theory
containing a spacetime dependent mass. 

It is not possible to give a similarly explicit construction of 
time ordered products satisfying (t1)--(t8) with more than one factor. 
Using the ideas of ``causal perturbation theory'' (see e.g. \cite{s})
one can, however, give an inductive construction of the time ordered
products so that (t1)--(t8) are satisfied which is based upon the 
above construction of the Wick powers (i.e., time ordered products with one factor). 
These constructions are described in detail in~\cite{bf, df2, hw3} (see expecially \cite{hw3} for the proof 
that scaling property (t5) can be satisfied), and we will therefore only
sketch the key steps and ideas going into this inductive construction, referring the reader 
to the references for details.

The main idea behind the inductive construction is that the 
causal factorization property expressing the temporal ordering of the 
factors in the time ordered product already defines  
time ordered products with the desired properties for non-coinciding spacetime points once the
Wick powers are known. Namely, if e.g.
$\supp f_1$ is before $\supp f_2$, $\supp f_2$ is before $\supp f_3$ etc., then  
the causal factorization property tells us that we must have
\ben
\label{21}
T(\prod_i f_i \Phi_i) = \, \Phi_1(f_1)  \cdots \Phi_n(f_n). 
\een
Since the Wick powers on the right side have already been constructed, 
we may take this relation as the definition of the time ordered products for 
testfunctions\footnote{Note that the time ordered products can 
be viewed as multi linear maps $\times^n \D(\mr^d) \to \W$
for a fixed choice of fields $\Phi_1, \dots, \Phi_n$.} 
$F = \otimes_i^n f_i$ whose support has no intersection
with any of the ``partial diagonals'' 
\ben 
\label{22}
\quad D_I = \{(x_1, \dots, x_{n}) \in \times^{n} \mr^d \mid x_i = x_j  \quad \forall i, j \in I\}, 
\quad I \subset \{1, \dots, n\},    
\een
in the product manifold $\times^n \mr^d$, because one can decompose such $F$ into 
contributions whose supports are temporally ordered via a partition of 
unity~\cite{bf}. The causal factorization property alone therefore already defines  
the time ordered products as $\W$-valued distributions, denoted $T^0$,
on the space $\times^n \mr^d$, minus the union $\cup_I D_I$ of all partial diagonals, 
and it can furthermore be seen that these objects have the desired properties (t1)--(t8)
on that domain. In order to define the time ordered products as distributions 
on all of $\times^n \mr^d$, one has to construct a suitable extension $T$ of $T^0$ to
a distribution defined on all of $\times^n \mr^d$ in such 
a way that (t1)--(t8) are preserved in the extension process. This step corresponds to the 
usual ``renormalization'' step in other approaches and is the hard part of the analysis. 
Actually, we can even assume that $T^0$ is already defined everywhere apart from the 
{\em total} diagonal $D_n = \{x_i \neq x_j \,\, \forall i, j\}$, since one can construct 
the extension $T$ inductively in the number of factors. Having constructed these for up 
to less or equal than $n-1$ factors then leaves the time ordered products with $n$ factors undetermined
only on the total diagonal. 

A key simplification for the extension problem occurs because the commutator condition (inductively known 
to hold for $T^0$) can be shown to be equivalent to the following
``Wick-expansion'' for $T^0$, 
\begin{multline}
\label{wickexpt}
T^0 \left( \prod_{i=1}^n \Phi_i(x_i) \right) = 
\sum_{\alpha_1, \alpha_2, \dots} \frac{1}{\alpha_1! \dots \alpha_n!} \times\\
\tau^0\left[ \delta^{\alpha_1} \Phi_1 \otimes \dots \otimes \delta^{\alpha_n} \Phi_n
\right](x_1, \dots, x_n)
: \prod_{i=1}^n \prod_j [(\partial)^j \phi(x_i)]^{\alpha_{ij}} :_H. 
\end{multline}
Here, the $\tau^0[\otimes_i \Psi_i]$ are c-number distributions on $\times^n \mr^d \setminus D_n$ 
[in fact equal to the expectation value of the time ordered product $T^0(\prod_i \Psi_i)$] depending 
in addition upon an arbitrary collection of fields $\Psi_1 \otimes \dots \otimes \Psi_n \in \otimes^n \V$, 
each $\alpha_j$ is a multi index and we are using the notation
\ben
\label{dledef}
\delta^\alpha \Phi = \left\{ \prod_j \left( \frac{\partial}{\partial [(\partial)^j \phi]} 
\right)^{\alpha_j} \right\} \Phi \in \V
\een
as well as $\alpha! = \prod_j \alpha_j!$ for multi indices\footnote{We are also suppressing tensor
indices in eqs.~\eqref{wickexpt} and~\eqref{dledef}. For example, the notation $(\partial)^j \phi$ is 
a shorthand for $\partial_{(\mu_1} \dots \partial_{\mu_j)}\phi$}. The notation $: \, \, :_H$ 
stands for the ``$H$-normal ordered products'', defined by 
\ben
\label{Hnormal}
: \prod_{i=1}^k \phi(x_i) :_H \, = \frac{\delta^k}{i^k \delta f(x_1) \dots \delta f(x_k)} 
e^{i \phi(f) + \frac{1}{2} H^{(m)}(f, f)} \Bigg|_{f=0}.
\een
The key point about the 
Wick expansion is that it reduces the problem of extending the algebra valued $T^0$
to the problem of extending the c-number distributions $\tau^0$. Since we want the 
extensions $T$ to satisfy (t1)--(t8), we also want the extensions $\tau$ of the $\tau^0$ 
to satisfy a number of corresponding properties: First, the wave front set condition 
on the $T$ correspond to the requirement that $\WF(\tau) \subset {\mathcal C}_T$, where
the set ${\mathcal C}_{T}$ was defined above in eq.~\eqref{gamtdef}. Second, since
the $T$ are required to be Poincare invariant, also the extension $\tau$ must be Poincare
invariant. Finally, since the $T$ are supposed to have an almost homogeneous scaling 
behavior under a rescaling $x \to \lambda x$ (and a simultaneous rescaling $m \to \lambda^{-1}
m$ of the mass), the $\tau$ must have the scaling behavior
\ben
\left(\frac{\partial}{\partial \log \lambda} \right)^k
\Big\{ \lambda^{D}
\tau^{(\lambda^{-1} m)}(\lambda x_1, \dots, \lambda x_n) 
\Big\} = 0,
\een 
for some $k$, where $D$ is the sum of the mass dimensions of the fields $\Psi_i$ on which $\tau$ depends, and 
where we are indicating explicitly the dependence of $\tau$ upon the mass parameter\footnote{The unitarity 
condition on the $T$ also implies a certain reality condition on the $\tau$, which however is rather 
easy to satisfy in the present context.}. 
By induction, these properties are already known for $\tau^0$ (i.e., off the total diagonal $D_n$), so
the question is only whether they can also be satisfied in the extension process. To reduce this remaining 
extension problem to a simpler task, one shows~\cite{hw3} that it is possible to expand the $\tau^0$ in 
terms of the mass parameter $m$ in a ``scaling expansion'' of the form
\ben
\tau^{(m) \, 0} = \sum_{k=n}^j m^{2k} \cdot u^0_k + r_j^0, 
\een
where the $u_j^0$ are Poincare invariant 
distributions (independent of $m$) that scale almost homogeneously 
under a rescaling of the spacetime coordinates,
\ben
\left(\frac{\partial}{\partial \log \lambda} \right)^k
\Big\{ \lambda^{D-2k}
u_k^0(\lambda x_1, \dots, \lambda x_n) 
\Big\} = 0, 
\een
with $\WF(u^0_k) \subset {\mathcal C}_T$, 
and where the remainder $r_j^0$ is a distribution with $\WF(r^0_j) \subset {\mathcal C}_T$, smooth in $m$, 
whose scaling degree~\cite{bf} can be made arbitrarily low by carrying out the expansion to 
sufficiently large order $j$. The idea is now to construct the desired extension $\tau$ by constructing 
separately suitable extensions of $u^0_k$ and $r^0_j$. Actually, since the remainder has a sufficiently low 
scaling degree, it extends by continuity to a unique distribution $r_j$, and that extension is 
seen to be automatically Poincare invariant, have wave front set $\WF(r_j) \subset {\mathcal C}_T$, 
and have an almost homogeneous scaling behavior under a rescaling of the coordinates and the mass 
parameter. The distributions $u_k^0$, on the other hand, do not extend by continuity, but one can 
construct the desired extension as follows~\cite{hw3}: One first constructs, by the methods originally
due to Epstein and Glaser and described e.g. in~\cite{s}, an arbitrary extension that is translationally
invariant and has the same scaling degree as the unextended distribution. That extension then also 
satisfies the wave front set condition~\cite{bf}, but it will not, in general, yield a distribution
with an almost homogeneous scaling behavior (i.e., homogeneous scaling up only to logarithmic terms), 
nor will it be Lorentz invariant. The point is, however, that this preliminary extension can always be 
modified, if necessary, so as to restore the almost homogeneous scaling behavior and Lorentz invariance
(while at the same time keeping the wave front set property and translational invariance), see 
lemma 4.1 of ref.~\cite{hw3}.
This accomplishes the desired extension of the $\tau^0$, and thereby establishes
the existence of a prescription $T$ for time ordered products satisfying (t1)--(t8).  

We emphasize that 
the above list of properties (t1)--(t8) does not determine the Wick powers and time ordered products
uniquely (for the time ordered products, this non-uniqueness arises because the extension 
process is not unique). The non-uniqueness corresponds to the usual
``finite renormalization ambiguities''.  Their form  
is severly restricted by the properties (t1)--(t8) 
and is described by the ``renormalization group'', see section~5. 

\section{Algebraic construction of the field observables as polynomials in $1/N$}

In this section, we generalize the algebraic construction of the 
field observables from a single scalar field 
to a multiplet of scalar fields, and we will show that the number of field components 
can be viewed as a free parameter that can be taken to infinity in a meaningful way
on the algebraic level. The model that we want to consider is described by the classical action
\ben
\label{23}
S = \int \tr (\partial^\mu \phi 
\partial_\mu \phi + m^2 \phi^2) \, d^d x, 
\een
where $\phi = \{\phi_{ij'}\}$ is now a field taking values in the 
hermitian $N\times N$ matrices, and where 
``$\tr$'' denotes the trace, with no implicit normalization factors. 
More precisely, we should think of the field as taking values in the 
${\bf N} \otimes \bar {\bf N}$ representation\footnote{We are putting a prime on 
the indices associated with the tensor factor transforming under $\bar {\bf N}$ in the spirit of 
van der Waerden's notation.} of the group $U(N)$, 
the trace being given by $\tr \phi^a = 
\sum \phi_{ij'} \delta^{j'k} \phi_{kl'} \delta^{l'm} \dots \phi_{mn'} \delta^{n'i}$ 
in terms of the invariant tensor $\delta_{ij'}$ in ${\bf N} \otimes \bar {\bf N}$. 

For an arbitrary, but fixed $N$, we begin by constructing a minimal algebra algebra
of observables corresponding to the action~\eqref{23} in a similar way as described 
in the previous section for the case of a single field. The minimal algebra is now generated by a unit 
and finite sums of products of smeared field components, $\phi_{ij'}(f)$, 
where $f$ runs through all compactly supported testfunctions, and where the ``color indices''
$i$ and $j'$ run from 1 to $N$. 
The relations in the case of general $N$ differ from relations 1)--4) for a single field only in 
that the hermiticity and commutation relations now read
\begin{enumerate}
\item[$3_N$.] (hermiticity)
$\phi_{ij'}(f)^* = \phi_{j'i}(\bar f)$
\item[$4_N$.] (commutator) $[\phi_{ij'}(f), \phi_{kl'}(h)] = 
i\delta_{il'}\delta_{kj'} \Delta(f,h) \cdot \myid$, 
\end{enumerate}
where $\Delta$ is the advanced minus retarded propagator of a single 
Klein-Gordon field. We construct an enarged algebra, $\W_N$, 
by passing to a new set of generators of the form~\eqref{3} and by allowing these generators 
to be smeared with suitable distributions, i.e., 
$\W_N$ is spanned by expressions of the 
form 
\ben
\label{25}
A = \int : \prod_k^a \phi_{i_k j_k{}'} (x_k) : t(x_1, \dots, x_a) \prod_{k=1}^a d^d x_k, 
\een
where $t \in \E'_a$ and where we are using the usual informal integral notation for 
distributions. The product of these quantities can again be expressed
in a form that is similar to~\eqref{5}.
Since the real components of the field $\phi_{ij'}$ are 
not coupled to each other, the enlarged algebra $\W_N$ is isomorphic to the 
tensor product of the corresponding algebra $\W_1$ for each independent real
component of the field as defined in the previous section, 
\ben
\label{24}
\W_N \cong \bigotimes^{N^2} \, \W_1, 
\een
$N^2$ being the number of independent real components of the field 
$\phi_{ij'}$. 

The transformation $\phi_{ij'} \to 
U_i{}^k \bar U_{j'}{}^{l'} \phi_{kl'}$ leaves the classical action functional~\eqref{23}
invariant for any unitary matrix $U \in U(N)$. This invariance property
is expressed on the algebraic level by a corresponding action of the group $U(N)$ on 
the algebras  $\W_N$ via a group of *-automorphisms $\alpha_U$. 
We are interested in the subalgebra 
\ben
\label{26}
\W^\inv_N = \{ A \in \W_N \mid \alpha_U(A) = A \quad \forall U \in U(N) \}   
\een
of ``gauge invariant'' elements, i.e., 
the subalgebra of $\W_N$ consisting of those 
elements that are invariant under this automorphic action of the group $U(N)$. 
It is not difficult to convince oneself that, as a vector space,  
$\W^\inv_N$ is given by
\ben
\label{27}
\W^\inv_N = {\rm span} \{W_a(t) \mid t \in \E'_{|a|} \}, 
\een
where 
\ben
\label{28}
W_{a}(t) = \frac{1}{N^{|a|/2}} \int
\, : 
\tr \left( \prod_{i_1 \in I_1} \phi(x_{i_1}) \right)
\cdots \tr \left( \prod_{i_T \in I_T} \phi(x_{i_T}) \right): t(x_1, \dots, x_{|a|}) \prod_{i=1}^{|a|} d^d x_i.    
\een 
Here, $t \in \E'_{|a|}$, the symbol 
$a$ now stands for a multi index, $a = (a_1, \dots, a_T)$, and we are using the 
usual multi index notation
\ben
\label{29}
|a| = \sum_i^T a_i.
\een
The $I_j$ are mutually disjoint 
index sets containing each $a_j$ elements such that $\cup_j I_j = \{1, \dots, |a|\}$. 
For later convenience, we have also incorporated an overall normalization factor into
our definition of the generators $W_a(t)$.
The generators are symmetric under exchange 
of the arguments within each trace and under exchange of the traces\footnote{Another
way of saying this is that the $W_a$ really act on distributions $t$ with 
these symmetry properties.}. Furthermore, they
satisfy a wave equation and hermiticity condition completely 
analogous to the ones in the scalar case, 
\ben
\label{30}
W_{a}(\bar t) = W_{a}(t)^*, 
\quad W_{a}([1 \otimes \cdots (\partial^\mu 
\partial_\mu - m^2)\otimes \cdots 1]t) = 0,   
\een
where the Klein-Gordon operator acts on any of the arguments of $t$.
  
Since $\W^\inv_N$ is a subalgebra of $\W_N$ (i.e., closed 
under multiplication), the product of two such 
generators can again be expressed as a linear combination of these 
generators. In order to determine the precise 
form of this linear combination, one has to take care of the 
color indices and the structure of the traces. 
Since the traces in the generators~\eqref{28} imply that there are ``closed loops'' of 
contractions of the color indices, there will also appear similar 
closed loops of index contractions in the formula for the product of 
two generators. Such closed index loops will 
give rise to combinatorical factors involving $N$. We are ultimately 
interested in taking the limit when $N$ goes to infinity, so we must study
the precise form of this $N$-dependence. 

Since the $N$-dependence arises solely from the index structure 
and not from the spacetime dependence of the propagators, 
it is sufficient for this purpose to study the ``matrix model'' given 
by the zero-dimensional version of the action functional~\eqref{23}, 
\ben
\label{31}
S_{\rm matrix} = m^2 \tr M^2, 
\een
where we have put $M = \phi$ in this case in order to emphasize the fact that we
are dealing now with hermitian 
$N \times N$ matrices $M = \{M_{ij'}\}$ with no dependence upon the 
spacetime point (this action does not, of course, describe a quantum field theory). 
By analogy with eq.~\eqref{3}, we define the ``normal ordered product'' of $k$
matrix entries to be the function
\ben
\label{32}
: M_{i_1 j_1{}'} \cdots M_{i_k j_k{}'} : \, = 
(-i)^k \frac{\partial^k}{\partial J^{i_1 j_1{}'} \cdots \partial J^{i_k j_k{}'}}
\e^{iJ \cdot M + \frac{1}{m^2} J^2 } \bigg|_{J = 0}  
\een
of the matrix entries where we have put $M \cdot J = \sum_{ij} M_{ij'} J^{ij'}$.
For the first values of $k$, the definition yields $:M_{ij'}: \, = M_{ij'}$, and 
$: M_{ij'} M_{kl'} : \, = M_{ij'} M_{kl'} - \delta_{il'} \delta_{kj'}/m^2$, etc. 
The (commutative) product of normal ordered polynomials $:Q(\{M_{ij'}\}):$ 
can be expressed in terms of normal ordered polynomials via the 
following version of Wick's theorem:
\ben
\label{33}
:Q_1: \cdots :Q_r: \, = \, :\e^{iM \cdot \partial/\partial {J^{}}} : 
\, \langle :\e^{-iJ^{} \cdot \partial/\partial M} Q_1: \cdots 
:\e^{-iJ^{} \cdot \partial/\partial M} Q_r: \rangle_{\rm matrix} \bigg|_{J^{}=0},  
\een
where we have introduced the ``correlation functions''
\ben
\label{34}
\langle :Q_1: \cdots :Q_r: \rangle_{\rm matrix} \equiv
{\mathcal N} \int dM  \, :Q_1: \cdots :Q_r: \, e^{-S_{\rm matrix}}  
\een
which we normalize so that $\langle 1 \rangle_{\rm matrix} = 1$. It follows from these 
definitions that we always have $\langle :Q: \rangle_{\rm matrix} = 0$. 
The correlation functions 
can be written as a sum of contributions associated with Feynman diagrams. Each such 
diagram consists of $r$ vertices that are connected by ``propagators'' 
\ben
\label{35}
\langle M_{ij'} M_{kl'} \rangle_{\rm matrix} = \frac{1}{m^2}\delta_{il'} \delta_{kj'}, 
\een
which are represented by oriented double lines, the arrow always going from the 
primed to the unprimed index. 

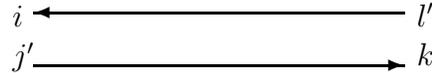
\begin{center}
\begin{picture}(200,80)
\thicklines
\put(30,30){\vector(1,0){140}}
\put(170,50){\vector(-1,0){140}}
\put(22,30){$j'$}
\put(175,30){$k$}
\put(22,45){$i$}
\put(175,45){$l'$}
\end{picture}
\end{center}
\noindent
The structure of $i$-th vertix is 
determined by the form of the polynomial $Q_i$. 

The space of gauge invariant polynomials in $M_{ij'}$ is spanned by 
functionals of the form\footnote{Note that these polynomials are not 
linearly independent at finite $N$. For example, for $N=2$ there holds
the relation $\tr M^3 - \frac{3}{2} \tr M \tr M^2 + \frac{1}{2} (\tr M)^3 = 0$. 
A set of linearly independent polynomials can 
be obtained using so-called ``Schur-polynomials''.}
\ben
\label{36}
W_{a} = \frac{1}{N^{|a|/2}} \, : \tr M^{a_1} \cdots \tr M^{a_T}:. 
\een 
which is the 0-dimensional analogue of expression~\eqref{28}, with the only
difference that there is no dependence on the smearing distribution $t$ since we 
are in 0 spacetime dimensions. Since these multi trace observables span 
the space of all polynomial $U(N)$-invariant function of the 
matrix entries, we already know that the product of 
$W_a$ with $W_b$
can again be written as a linear combination of such
observables. We are interested in the dependence of the 
coefficients in this linear combination on $N$. 
We calculate the product $W_a \cdot W_b$ via Wick's formula~\eqref{33}, 
and we organize the resulting sum of expressions in terms of the following Feynman graphs. 
From the $T$ traces in $W_a$, there will be $T$ $a$-vertices with $a_j$ legs each (we think of the 
legs as carring a number), where $j = 1, \dots, T$. We draw the legs as double lines. As an 
example, let $a = (a_1, a_2) = (3, 3)$, so that $W_{(3, 3)} = 1/N^{3} : \tr M^3 \tr M^3 :$. In this
case, we have 2 $a$-vertices with with $3$ lines, each corresponding to one trace 
with 3 factors of $M$. Each such vertex looks therefore as follows:

\begin{center}
\begin{fmffile}{graph6}
\begin{fmfgraph*}(100,80) 
\fmfpen{thick}
\fmfleft{i1,i2}
\fmfright{o1}
\fmflabel{$ij'$}{i1}
\fmflabel{$jk'$}{i2}
\fmflabel{$ki'$}{o1}
\fmf{dbl_plain}{i1,w1,i2}
\fmf{dbl_plain}{o1,w1}
\end{fmfgraph*}
\end{fmffile}
\end{center}

The double lines should also be equipped with orientations that are compatible at the vertex, although
we have not drawn this here. From the $S$ traces in $W_b$ 
there are similarly $S$ $b$-vertices with $b_j$ legs each, where $j = 1, \dots, S$.
We consider graphs obtained by joining $a$-vertices with $b$-vertices by a double line representing a 
matrix propagator~\eqref{35}, but we do not allow any $a-a$ or $b-b$ connections 
(such connections have already been taken care of by the normal ordering prescription used
in the definition of $W_{a}$ respectively $W_{b}$).  
We finally attach an ``external current'' $M_{ij'}$ to every leg of an $a$-vertex
or $b$-vertex that is not connected by a propagator. An example of a 
Feynman graph resulting from this procedure occurring in the product $W_{(a_1, a_2)} \cdot 
W_{(b_1, b_2)}$ with $a_1 = a_2 = b_1 = b_2 = 3$ is drawn in the following picture.

\begin{center}
\vspace{1cm}
\begin{fmffile}{graph4}
\begin{fmfgraph*}(120,100) 
\fmfpen{thick}
\fmfleft{i1,i2}
\fmfright{o1,o2}
\fmfv{label=$a_1=3$,label.angle=180}{w1}
\fmfv{label=$b_1=3$,label.angle=180}{w2}
\fmfv{label=$b_2=3$,label.angle=0}{w3}
\fmfv{label=$a_2=3$,label.angle=0}{w4}
\fmf{dbl_plain}{i1,w1,w3,o1}
\fmf{dbl_plain}{o2,w4,w2,i2}
\fmf{dbl_plain}{w1,w2}
\fmf{dbl_plain}{w3,w4}
\fmflabel{$M_{ij'}$}{i1}
\fmflabel{$M_{jk'}$}{i2}
\fmflabel{$M_{li'}$}{o1}
\fmflabel{$M_{kl'}$}{o2}
\fmf{plain}{i1,i1}
\fmf{plain}{i2,i2}
\fmf{plain}{o1,o1}
\fmf{plain}{o2,o2}
\end{fmfgraph*}
\end{fmffile}
\end{center}

\vspace{1cm}

The resulting structure will consist of a number of closed loops obtained
by following the lines (including loops that run through external currents). 
There will be, in general, 3 kinds of loops: 
(1) Degenerate loops around a single vertex that has only
external currents but no propagators attached to it. Let the number 
of such loops (i.e., isolated vertices) be $D$.
(2) Loops that 
contain at least one external current and at least one propagator line. Let the number of these
surfaces be $J$. (3) Loops that contain no external current. Let $I$
be the number of these loops. [Thus, in the above example graph, we have $D=0$, $I=1$ (corresponding
to the inner square-shaped loop) and $J=1$ (the loop running around the square passing through the 
4 external currents).]

Following a set of ideas by `t Hooft~\cite{th}, we consider the big (in general 
multiply connected) closed 2-dimensional
surface $\S$ obtained by capping off the loops of type (2) and (3) with little 
surfaces (we would obtain a sphere in the above example.)
The total number $F$ of little surfaces in $\S$ is consequently given by
\ben
\label{37}
F = I + J.
\een
Let us label the loops containing currents by $j=1, \dots, D+J$, and let $c_j$ be the 
the number of currents in the corresponding loop.
By construction, the number of edges, $P$, of the surface $S$ is related 
to $a$, $b$, and $c$ by
\ben
\label{38}
2P = |a| + |b| - |c|,   
\een
where we are using the same multi index notation as above.
The number of vertices, $V$, in $\S$ is 
\ben
\label{39}
V = T + S - D, 
\een
i.e., is equal to the total number of traces $T$ and $S$ in 
$W_{a}$ and $W_{b}$ minus $D$, the number 
of vertices that are not connected to any other vertex. We  
apply the well-known theorem by Euler to the surface $\S$ which tells
us that 
\ben
\label{40}
F - P + V = \sum_k (2-2H_k), 
\een 
where $H_k$ is the genus of the $k$-th disconnected
component of $\S$. The little surfaces in $\S$ each 
carry an orientation induced by the direction of the enclosing 
index loops, and these give rise to an orientation on each of 
the connected components of the big surface $\S$. 
An oriented 2-dimensional surface always has $H_k \ge 0$, 
and $H_k$ is equal to the number of handles of the corresponding 
connected component in that case.

Let us analyze the contributions to the product $W_{a} \cdot W_{b}$ 
associated with a given 
graph. From the $P$ double lines of the graph, there will be a contribution
\ben
\label{41}
\prod_{\text{lines $(k,l)$}} \frac{1}{m^2}, 
\een
associated with the double line propagators. From the closed loops of the kind (3) in the graph 
there will be a factor
\ben
\label{42}
N^I = N^{\sum(2-2H_k) + (|a|+|b|-|c|)/2 - V - J}
\een
because each of the $I$ such closed index loops 
gives rise to a closed loop of index contractions of Kronecker deltas, 
$N = \sum \delta_i{}^i$. Finally, there will be 
a contribution 
\ben
\label{43}
: \tr M^{c_1} \cdots \tr M^{c_{J+D}} :
\een
corresponding to the external currents in 
$J+D$ closed loops of the kind (1) and (2) containing $c_i$ external 
currents each. Taking into account the 
normalization factors of $N^{-|a|/2}$ respectively $N^{-|b|/2}$
associated with $W_{a}$ respectively $W_{b}$,  
and letting $V_k$ be the number of vertices in the $k$-th 
connected component of $\S$, we therefore find
\ben
\label{44}
W_{a}\cdot W_{b} 
= \sum_{\rm graphs} 
(1/N)^{J + \sum H_k + \sum(V_k-2)}
\prod_{\text{lines $(k,l)$}} \frac{1}{m^2} 
\cdot W_{c},  
\een
where the sum is over all distinct Feynman graphs\footnote{Note that we think 
of the legs of $a$- and $b$-vertices as numbered, and so a graph is understood 
here as a graph carrying the corresponding numberings. Topologically identical 
graphs with distinct numberings of the legs count as different in the above sum, 
as well as similar sums below.}.

We can easily generalize these considerations to calculate the 
product of $W_{a}(f_1\otimes \dots \otimes f_{|a|})$ with 
$W_{b}(h_1\otimes \dots\otimes h_{|b|})$ in the case when the 
dimension of the spacetime is non zero. In this case, the legs of each 
$a$-vertex are associated with the smearing functions $f_j$ 
appearing in the corresponding trace in eq.~\eqref{28}, 
and the legs of every $b$-vertex are likewise associated with a smearing functions
$h_k$. Every matrix propagator connecting such an $a$ and $b$ vertex
then gets replaced by $\Delta_+$, 
\ben
\label{45}
\frac{1}{m^2} \to \Delta_+(f_j, h_k).  
\een
Furthermore, in a given graph, the $J+D$ index loops with 
currents now correspond to a contribution of the form\footnote{To simplify, 
we are assuming here that $\Delta_+$ is given by the Wightman 2-point function, 
see eq.~\eqref{10}.}
\ben
\label{46}
: \tr \left( \prod^{c_1} \phi(j_i) \right) \cdots 
\tr \left( \prod^{c_{J+D}} \phi(j_k) \right) :, 
\een
where the $j_k \in \{f_k, h_k\}$ denotes the test function associated 
with the corresponding external current. With these replacements, we obtain 
the following formula
for the product of two generators $W_{a}(\otimes_i f_i)$ with 
$W_{b}(\otimes_i h_i)$ of the algebra $\W^\inv_N$:
\begin{multline}
\label{47}
W_{a}(f_1 \otimes \cdots \otimes f_{|a|}) \cdot W_{b}
(h_1 \otimes \cdots \otimes h_{|b|}) \\
= \sum_{\rm graphs} 
(1/N)^{J + \sum H_k + \sum(V_k-2)}
\prod_{\text{lines $(k,l)$}} \Delta_+(f_k, h_l) \\
\cdot W_{c}(j_1\otimes \cdots \otimes j_{|c|}).
\end{multline}
An entirely analogous formula is obtained if the test functions
$\otimes_i f_i$ and $\otimes_j h_j$ are replaced by arbitrary distributions
$t$ and $s$ in the spaces $\E'_{|a|}$ respectively $\E'_{|b|}$. 

The important thing to observe about relation eq.~\eqref{47}
is how the coefficients in the sum on the right side 
depend on $N$: The numbers $H_k$ (the 
number of handles of the $k$-th component of the 
surface $\S$ associated with the graph) and $J$ are always non-negative. 
The number $V_k - 2$
is also non-negative since the number of vertices
in each component, $V_k$, is by construction always greater or equal than 2.
Hence, we conclude that $1/N$ appears always with a non-negative power in 
the coefficients on the right side of eq.~\eqref{47}. 
Since these coefficients are essentially the ``structure constants'' of the 
algebra $\W^\inv_N$, it is therefore possible take the large $N$ limit
on the algebraic level. We now formalize this idea by constructing a new algebra which 
has essentially the same relations as the algebras $\W^\inv_N$, 
but which incorporates the important new point of view that $N$, or rather
\ben
\label{48}
\eps = \frac{1}{N}
\een
is not fixed, but is instead considered as a free expansion parameter that can 
range freely over the real numbers, including in particular $\eps = 0$.   
We will then show that this algebra contains elements corresponding
to Wick powers and their time ordered products. The construction of 
this new algebra therefore incorporates the $1/N$ expansion of the 
quantum field observables associated with the action~\eqref{23}, including in 
particular the large $N$ limit of the theory.

Consider the complex vector space $\X[\eps]$ 
consisting of formal power series expressions of the form
\ben
\label{49}
\sum_{j \ge 0} \eps^j W_{a_j}(t_j)
\een
in the ``dummy variable'' $\eps$, where the $a_j$ are 
multi-indices, and where the $t_j$ are taken from the 
space $\E'_{|a_j|}$ of distributions in $|a_j|$ spacetime 
variables defined in~\eqref{7}. We implement the second of relations eq.~\eqref{30} by 
viewing the symbols $W_{a_j}(t_j)$ as depending only on the 
equivalence class of $t_j$ in the quotient space 
${\mathcal J}_{|a_j|}$, where 
\ben
\label{50}
{\mathcal J}_n = \E'_n/
\{ (1 \otimes \cdots (\partial^\mu \partial_\mu - m^2) \otimes \cdots 1)s
\mid s \in {\mathcal E}'_n\}. 
\een
On the so defined complex vector space
$\X[\eps]$, we define a product by 
\ben
\label{51}
\left( \sum_{j \ge 0} \eps^j W_{a_j}(t_j) \right)
\left( \sum_{k \ge 0} \eps^k W_{b_k}(s_k) \right) = 
\sum_{r \ge 0} \eps^{r} \sum_{r=k+j} W_{a_j}(t_j) \cdot
W_{b_k}(s_k), 
\een
where the product $W_{a_j}(t_j) \cdot
W_{b_k}(s_k)$ is given by formula eq.~\eqref{47} 
(with $1/N$ replaced by $\eps$ in this formula), and we define a
*-operation on $\X[\eps]$ by 
\ben
\label{52}
\left( \sum_{j\ge 0} \eps^j W_{a_j}(t_j) \right)^* 
= \sum_{j \ge 0} \eps^j W_{a_j}(\bar t_j).
\een 
\begin{prop}
The product formula~\eqref{51}
and the formula~\eqref{52} for the *-operation makes $\X[\eps]$ into an 
(associative) *-algebra with unit (given by $\myid \equiv W_0$). 
\end{prop} 
\begin{proof}
We need to check that the product formula~\eqref{51} defines an associative 
product, and that the formula~\eqref{52} for the *-operation is compatible 
with this product in the usual sense. 
For associativity, we consider the associator of generators
\ben
\label{53}
A(\eps) = W_a(r) \cdot (W_b(s) \cdot W_c(t)) - (W_a(r) \cdot W_b(s)) \cdot W_c(t),  
\een
which we evaluate using the product formula in the order specified by 
the brackets. The resulting expression can be written as a finite sum 
of terms of the form $\sum Q_j(\eps) W_{d_j}(u_j)$, 
where the $Q_j(\eps)$ are polynomials in $\eps$, and where the $u_j$ 
are linearly independent. But we already know $A(\eps) = 0$ for 
$\eps = 1, 1/2, 1/3,$ etc., since the algebras $\W_N^\inv, N=1, 2, 3,$ etc. are associative. 
Therefore, the $Q_j(\eps)$ must vanish for these values of $\eps$. Since a polynomial 
vanishes identically if it vanishes when evaluated on an infinite set of distinct real numbers, 
it follows that the $Q_j$ vanish identically, proving that 
$A(\eps)=0$ as a power series in $\eps$. The consistency of 
the *-operation is proved similarly.
\end{proof}

The construction of the algebra $\X[\eps]$ completes our desired algebraic formulation of the $1/N$ expansion 
of the field theory associated with the free action~\eqref{23}. Our construction of $\X[\eps]$
depends on a particular choice of the distribution $\Delta_+$, but 
different choices again give rise to isomorphic algebras, showing that, as an abstract algebra, 
$\X[\eps]$ is independent of this choice. Indeed, let $\Delta_+'$ be another bidistribution
whose antisymmetric part is $(i/2) \Delta$ which satisfies the wave 
equation and has wave front set $\WF(\Delta_+')$ of Hadamard form,
and let $\X'[\eps]$ be the corresponding algebra constructed 
from $\Delta_+'$ with generators $W'_a(t)$. Then the desired 
*-isomorphism from $\X[\eps] \to \X'[\eps]$ is given by
\ben
\label{54}
W_a(t) \to \sum_{\rm graphs} 
\eps^{J + \sum H_k + \sum(V_k-2)}
\cdot W_{b}'\left( \langle \bigotimes_{\text{lines}} F, t \rangle \right). 
\een
Here $F$ is the smooth function given by $\Delta_+ - \Delta_+'$ and a 
graph notation as in eq.~\eqref{47} has been used: The sum is over all 
graphs obtained by writing down the $T$ vertices corresponding to the 
$T$ traces in $W_a(t), a = (a_1, \dots, a_T)$, by contracting
some legs with ``propagators'', and by attaching ``external currents''
to others. If $x_i$ is the point associated with the $i$-th
leg in a graph with $n$ propagator lines ($2n \le |a|$), then 
\ben
\label{55}
\langle \bigotimes_{\text{lines}} F, t \rangle(x_1, \dots, 
x_{|a|-2n}) = 
\int t(x_1, \dots, x_{|a|}) 
\prod_{\text{lines $(i,j)$}} F(x_i, x_j) \, \prod_{\text{legs $i$}} d^dx_i, 
\een
where the second product is over all legs which have a propagator attached 
to them. The numbers $H_k$ and $V_k$ are the number of handles respectively 
the number of vertices ($\ge 2$) in the $k$-th disconnected component 
of the surface associated with the graph. $J$ is the number of closed 
index loops associated with the graph which contain currents, and each 
such loop with $b_i$ currents 
corresponds to a trace in $W'_b$, where $b = (b_1, b_2, \dots, b_J)$. 
We note that this implies 
in particular that only positive powers of $\eps$ appear
in eq.~\eqref{54}, which is necessary in order for the right side to be 
an element in $\X[\eps]$. 

\medskip

We finally show that $\X[\eps]$ contains observables corresponding to 
the suitably normalized smeared gauge invariant Wick powers and their time 
ordered products. To have a reasonably compact notation for these objects, 
let us introduce the vector space of all formal gauge 
invariant expressions in the field $\phi$ and its derivatives, 
\ben
\label{56}
\V^\inv = \text{span} 
\left\{ \Phi = \prod_i \tr \left( \prod \partial_{\mu_1} \cdots \partial_{\mu_k} \phi \right) \right\}.
\een
If $\Phi$ is a monomial in $\V^\inv$, we denote by $|\Phi|$ the number of 
free field factors $\phi$ in the formal expression for $\Phi$, for example $|\Phi|=6$ for the 
field $\Phi = \tr \phi^4 \tr \phi^2$. 
For a fixed $N$, the gauge invariant Wick powers are viewed as linear maps
\ben
\label{57}
\D(\mr^d, \V^\inv) \to \W^\inv_N, \quad f\Phi \to \Phi(f).  
\een
Likewise, the gauge invariant time ordered products are viewed as multi linear maps 
\ben
\label{59}
T: \times^n \D(\mr^d; \V^\inv) \to \W^\inv_N, \quad
(f_1 \Phi_1, \dots, f_n \Phi_n) \to T(f_1 \Phi_1 \cdots f_n \Phi_n). 
\een  
The Wick powers are identified with the time ordered products with a single factor.
In the previous section, we demonstrated that, in the scalar case ($N=1$), the 
Wick powers and time ordered products can be constructed so as to satisfy a 
number of properties that we labelled (t1)--(t8). It is clear that these constructions can be generalized 
straightforwardly also to the case of a multiplet of scalar fields in the adjoint representation
of $U(N)$ (with $N$ arbitrary but fixed) and thereby yield time ordered products with 
properties completely analogous to the properties (t1)--(t8) stated above for the scalar case. 
We would now like to investigate the dependence upon $N$ of these objects and show that, 
if the time ordered products are normalized by suitable powers of $1/N$, these can be 
viewed as elements of $\X[\eps]$, i.e., that they can be expressed as a linear 
combination of $W_a(t)$, with $t$ depending only on positive powers of $\eps = 1/N$. 
 
The Wick powers $\Phi(f)$ are constructed in the same way as in the scalar case, see eq.~\eqref{wickpowers}. 
The only difference is that we need to multiply the Wick powers by suitable normalization factors
depending upon $\eps = 1/N$ in order to get well-defined elements of $\X[\eps]$. Taking into account 
the normalization factor in the
definition of the generators $W_a$, eq.~\eqref{28}, one sees that 
\ben
\label{58}
\eps^{|\Phi|/2} \Phi(f) \in \X[\eps].  
\een
Given that the suitably normalized 
Wick powers eq.~\eqref{58} are elements in $\X[\eps]$, one naturally expects that 
their time ordered products are also elements in $\X[\eps]$, 
\ben
\label{59a}
\eps^{\sum |\Phi_i|/2} T \left( \prod_i f_i \Phi_i \right) \in \X[\eps].
\een
Now, if the testfunctions $f_i$ are temporally ordered, i.e., if for example
the support of $f_1$ is before $f_2$, the support of $f_2$ before $f_3$ etc., 
then the time ordered product factorizes into the ordinary algebra product 
$\eps^{|\Phi_1|/2} \Phi_1(f_1) \eps^{|\Phi_2|/2} \Phi_2(f_2) \dots$ in $\X[\eps]$, 
by the causal factorization property of the time ordered products. Therefore, 
since the normalized Wick powers have already been demonstrated to be 
elements in $\X[\eps]$, also their product is (because $\X[\eps]$ was shown to
be an algebra). Hence, one concludes by this 
arguments that if $F = \otimes_i f_i$ is supported away from the union of all 
partial diagonals $D_I$ in the product manifold $\times^n \mr^d$, then the 
corresponding time ordered product satisfies~\eqref{59a}. However, for arbitrarily
supported testfunctions $f_i$, the time ordered product does not factorize, and 
therefore does not correspond to the usual algebra product. In other words, 
whether eq.~\eqref{59a} is satisfied or not depends on the definition of the 
time ordered products on the partial diagonals $D_I$ (as a function of $N$). As we 
have described explicitly above in the scalar case, the definition of the time ordered products
on the diagonals is achieved by extending the time ordered products defined 
by causal factorization away from the diagonals in a suitable way. Therefore, in order 
that eq.~\eqref{59a} be satisfied, we must control the dependence upon $N$ of the constructions in 
the extension argument, or, said differently, we must control 
the way in which the time ordered products are renormalized as a function of $N$.

Acutally, we will now argue that one can construct time ordered products 
$T$ satisfying eq.~\eqref{59a} [in addition to (t1)--(t8)] from the scalar time
ordered products that were constructed in the previous section, so there is no 
need to repeat the extension step. The
arguments are purely combinatorical and very similar to the kinds of arguments
used before in the construction of the algebra $\X[\eps]$, so we will only sketch
them here. Also, to keep things as simple as possible, we will consider explicitly
only the case in which all the $\Phi_i$ contain only one trace and no derivatives, 
$\Phi_i = \tr \phi^{a_i}$. In order to describe our construction of $T(\prod_i \Phi_i)$, we 
begin by considering the coefficient distributions $\tau[\otimes_i \phi^{b_i}]$ 
occurring in the Wick expansion~\eqref{wickexpt} of the time ordered products $T(\prod \phi^{a_i})$
in the {\em scalar} theory $(N=1)$. As it is well-known, these can be decomposed into contributions
from individual Feynman graphs
\ben
\tau[\phi^{b_1} \otimes \cdots \otimes \phi^{b_n}] = \sum_{\text{graphs $\gamma$}} 
c^\gamma \cdot \tau^\gamma[\phi^{b_1} \otimes \cdots \otimes \phi^{b_n}]. 
\een
Here, the sum is over all distinct Feynman graphs $\gamma$ (in the scalar theory)
with $n$ vertices of valence $b_i$ each (and no external lines), 
and $c^\gamma$ is a combinatorical factor chosen so that 
$\tau^\gamma$ coincides with the distribution constructed by the usual Feynman rules
(where the latter are well-defined as distributions, i.e., away from the diagonals).
Consider now, in $\X[\eps]$, the product 
$\eps^{|\Phi_1|/2} \Phi_1(f_1) \eps^{|\Phi_2|/2} \Phi_2(f_2) \dots$. If one evaluates this
product successively using the product formula~\eqref{48}, then one sees that the result
is organized in terms the following double-line Feynman graphs $\Gamma$: Each such 
graph has $n$ vertices labelled by points $x_i$ of valence $a_i$ each (drawn as in the figure
at the bottom of p.~17). External lines ending
on a vertex $x_k$ carrying a color index pair $(ij')$ are associated with a factor of $\phi_{ij'}(x_k)$. 
Given a such a graph $\Gamma$, we form the surface ${\mathscr S}_\Gamma$ by capping off the 
closed index loops with little surfaces, i.e. the index loops that do not meet any of the 
factors $\phi_{ij'}(x_k)$. We do not cap off any of the index loops that meet one or more of the 
factors $\phi_{ij'}(x_k)$ along the way, and these will consequently correspond to holes in the 
surface. We let $J$ be the number of such holes and we let $h= 1, \dots, J$ be an index labelling the holes.  
We will say that $k \in h$ if the factor $\phi_{ij'}(x_k)$ is encountered when running around the 
index loop in the hole labelled by $h$. Finally, for a given 
double line graph $\Gamma$, let $\gamma$ be the single line graph obtained from $\Gamma$ by removing 
all the external lines, and by replacing all remaining double lines by single ones. Then, for 
$(x_1, \dots, x_n)$ such that $x_i \neq x_j$ for all $i,j$ --- i.e., away from all partial diagonals ---
we can rewrite $T(\prod \tr \phi^{a_i}(x_i))$
as follows:
\begin{multline}
\label{Nwickexp}
\eps^{\sum a_i/2} T \left( \prod_i \tr \phi^{a_i}(x_i) \right) = \sum_\Gamma
\eps^{J + \sum H_k + \sum (V_k-2)} \tau^\gamma(x_1, \dots, x_n) \times\\
: \prod_h \tr \left\{ \prod_{k \in h} \eps^{1/2} \phi(x_k) \right\} :_H, 
\end{multline}
where $V_k$ is the number of vertices in the $k$-th disconnected component of ${\mathscr S}_\Gamma$, and
$H_k$ the number of handles. The idea is now to {\em define} the time ordered product on the left 
side by the right side for arbitrary $(x_1, \dots, x_n)$, including configurarions on the partial 
diagonals. A similar definition can be given for operators $\Phi_i$ containing multiple traces 
or derivatives, the only difference being that the Feynman diagrams that are involved have to 
also incorporate the multiple traces.

The key point about our definition~\eqref{Nwickexp} is that we have now complete control over the dependence
upon $N$ of the time ordered products of gauge invariant elements: 
Since the expression in the second line of the above equation is an element of $\X[\eps]$
(after smearing), it follows that the so-defined $\eps^{\sum a_i/2} T ( \prod \tr \phi^{a_i}(x_i))$ is 
an element of $\X[\eps]$ (after smearing). Also, since the $\tau$ have been defined so that the corresponding time ordered 
products in the scalar theory (see eq.~\eqref{wickexpt}, with $\tau^0$ replaced by $\tau$ in that equation)
satisfy (t1)--(t8), it follows that the time ordered products at arbitrary $N$ defined by eq.~\eqref{Nwickexp} 
also satisfy these properties. A similar argument can be given when the operators $\Phi_i$ contain multiple
traces or derivatives. Thus, we have altogether shown that the algebra $\X[\eps]$ of formal power series 
in $\eps = 1/N$ contains the suitably normalized time ordered products of gauge invariant elements, and that 
these time ordered products can be defined so that they satisfy the analogs of (t1)--(t8).

\section{The interacting field theory}

In the previous sections, we constructed an algebra of observables
$\X[\eps]$ associated with the free field described by the action~\eqref{23}, 
whose elements are (finite) power series in the free parameter $\eps = 1/N$.   
This algebra contains, among others, the gauge invariant 
smeared Wick powers of the free field and their time ordered products. 
In the present section we will show how to construct from these building 
blocks the interacting field quantities as power series in $\eps$ 
and the self-coupling constant in an interacting quantum field theory 
with free part~\eqref{23} and gauge invariant interaction part,  
\ben
\label{60}
S = \int \tr (\partial^\mu \phi 
\partial_\mu \phi + m^2 \phi^2) + V(\phi) \, d^d x. 
\een
For definiteness we consider the self-interaction
\ben
\label{61}
V(\phi) = g\tr \phi^4 
\een
which will be treated perturbatively. 

We begin by constructing the 
perturbation series for the interacting fields for a given but {\em fixed} $N$. 
Let $K$ be a compact region in $d$-dimensional Minkowski spacetime, and
let $\theta$ be a smooth cutoff function with which is equal to 1 on $K$ and which vanishes 
outside a compact neighborhood of $K$. For the cutoff interation
$\theta(x) V$ and a given $N$, we define interacting fields by Bogoliubov's formula
\ben
\label{64}
\Phi_{\theta V}(f) \equiv \frac{\partial}{i\partial \lambda} S(\theta V)^{-1} S(\theta 
V + \lambda f\Phi)\bigg|_{\lambda=0},  
\een
where the {\em local $S$-matrices} appearing in the above equation are 
defined in terms of the time ordered products in the free theory by
\ben
\label{62}
S\left( g\sum_j f_j \Phi_j \right) = 
\sum_n \frac{(ig)^n}{n!} T \left( \prod^n \sum_j f_j\Phi_j \right), \quad \Phi_j \in \V^\inv. 
\een
Although each term in the power series defining the local 
$S$-matrix is a well defined element in the 
algebra $\W^\inv_N$, the infinite sum of these terms is not, since
this algebra by definition only contains finite sums of generators. We do 
not want to concern ourselves here with the problem of convergence of the 
perturbative series, so we will view the local $S$-matrix, and likewise the 
interacting quantum fields~\eqref{64}, simply as a formal power 
series in $g$ with coefficients in $\W^\inv_N$, that is, as elements of the vector space 
\ben
\label{63}
\W_N^\inv[g] = \left\{ \sum_{n = 0}^\infty A_n g^n \,\, \bigg| \,\, A_n \in \W_N^\inv 
\quad \forall n \right\}.  
\een
We make the space $\W_N^\inv[g]$ into 
a $*$-algebra by defining the product of two formal power series to be the formal power
series obtained by formally expanding out the product of the infinite sums, $(\sum_n A_n g^n)
\cdot (\sum_m B_m g^m) = \sum_k \sum_{m+n=k} (A_n \cdot B_m) g^k$, and by 
defining the *-operation to be $(\sum_n A_n g^n)^* = \sum_n A_n^* g^n$.

We now remove the cutoff $\theta$ on the algebraic level. For this, we first note that
the coefficients in the power series\footnote{This series is sometimes referred to as 
``Haag's series'', since it was first obtained in~\cite{ha}.} defining the 
interacting field~\eqref{64} with cutoff are in fact the so-called ``totally retarded products'', 
\ben
\label{65}
\Phi_{\theta V}(f) = \Phi(f) + \sum_{n \ge 1} \frac{(ig)^n}{n!} 
R(\underbrace{\theta \tr \phi^4 \dots \theta \tr \phi^4}_{\text{$n$
factors}}; f \Phi), 
\een 
each of which can in turn be written in terms of products of time ordered products. 
It can be shown that the retarded products vanish whenever the support of $\theta$ is not in the causal
past of the support of $f$. This makes it possible to define the interacting fields not only for 
compactly supported cutoff functions $\theta$, but more generally for cutoff functions with  
compact support only in the time direction, i.e., we can choose $K$ to be a time slice. 
Thus, when $\theta$ is supported in a time slice $K$, then the right side of eq.~\eqref{65} is 
still a well-defined element of $\W^\inv_N[g]$.

We next remove the restriction to interactions localized in a time slice. For this, we 
consider a sequence of cutoff functions $\{\theta_j\}$ which 
are 1 on time slices $\{K_j\}$ of increasing size, eventually covering 
all of Minkowksi spacetime in the limit as $j$ goes to infinity. It is tempting to 
try to define the interacting field without cutoff as the limit of the algebra elements
obtained by replacing the cutoff function $\theta$ in eq.~\eqref{64} by the members of the 
sequence $\{\theta_j\}$. This limit, provided it existed, would in effect correspond to defining
the interacting field in such a way that it coincides with the free ``in''-field in the asymtptoic 
past. However, it is well-known that such an ``in''-field will in general fail to make sense in the 
massless case due to infrared divergences. Moreover, it is clear that the local quantum fields 
in the interior of the spacetime should at any rate make sense no matter what the infrared behavior of the 
theory is. As we will see, these difficulties are successfully avoided if, instead of 
trying to fix the interacting fields as a suitable ``in''-field in the asymptotic past, 
we fix them in the interior of the spacetime. 

We now formalize this idea following~\cite{hw1} (which in turn is based on ideas of~\cite{bf}). 
For this, it is important that for any pair of cutoff functions $\theta, \theta'$ which 
are equal to 1 on a time slice $K$, there 
exists a unitary $U(\theta,\theta') \in \W^\inv[g]$ such that~\cite{bf}  
\ben
\label{66}
U(\theta,\theta') \cdot \Phi_{\theta V}(f) \cdot U(\theta,\theta')^{-1} = \Phi_{\theta' V}(f)
\een
for all testfunctions $f$ supported in $K$, and for all $\Phi$. 
These unitaries are in fact given by 
\ben
\label{67}
U(\theta, \theta') = S(\theta V)^{-1} S(h_- V),  
\een
where $h_-$ is equal to $\theta - \theta'$ in the causal past of $K$ and equal to 0 in the 
causal future of $K$. Equation~\eqref{66} shows in particular that, within $K$, the algebraic
relations between the interacting fields do not depend on one's choice 
of the cutoff function. From our sequence of cutoff functions $\{\theta_j\}$, 
we now define $u_1 = \myid$ and unitaries $u_j = U(\theta_j, \theta_{j-1})$ for $j>1$, and 
we set $U_j = u_1 \cdot u_2 \cdot \dots \cdot u_j$. 
We define the interacting field without cutoff to be 
\ben
\label{68}
\Phi_{V}(f) \equiv \lim_{j \to \infty} 
U_j \cdot \Phi_{\theta_j V}(f) \cdot U_j^{-1},  
\een
where $f$ is allowed to be an arbitrary testfunction of compact support.
In fact, using eqs.~\eqref{66} and~\eqref{67}, one 
can show (see prop.~3.1 of~\cite{hw1})  
that the sequence on the right side remains constant once $j$ is so large that $K_j$ 
contains the support of 
$f$, which implies that the right side is always a well-defined element of $\W_N^\inv[g]$. 
The unitaries $U_j$ in eq.~\eqref{68} implement the idea to ``keep the interacting field fixed 
in the interior of the slice $K_1$'', instead of keeping it fixed in the asymptotic past.
This completes our construction of the interacting fields without cutoff.

These constructions can be generalized to define time ordered products of interacting fields by first 
considering the corresponding quantities associated with the cutoff interaction $\theta(x) V$, 
\ben
\label{69}
T_{\theta V}(f_1 \Phi_1 \cdots f_n \Phi_n) 
\equiv \frac{\partial}{i^n\partial \lambda_1 \cdots \partial \lambda_n} 
S(\theta V)^{-1} S(\theta V + \sum_i \lambda_i f_i \Phi_i)\bigg|_{\lambda_i=0},  
\een
possessing a similar expansion in terms of retarded products, 
\ben
\label{70}
T_{\theta V}(\prod_i f_i \Phi_i) = 
T(\prod_i f_i\Phi_i)
+ \sum_{n \ge 1} \frac{(ig)^n}{n!} R(\underbrace{\theta \tr \phi^4 \dots \theta \tr \phi^4}_{\text{$n$
factors}}; \prod_i f_i \Phi_i).
\een
The corresponding time ordered products without cutoff, denoted $T_V(\prod f_i \Phi_i)$, are then
defined in the same way as the interacting Wick powers, see eq.~\eqref{68}. 
The latter are, of course, equal to the time ordered products with only one factor, 
\ben
\Phi_V(f) = T_V(f \Phi).
\een 

The definition of the interacting fields as elements of 
$\W^\inv_N[g]$ depends on the chosen sequence of time slices $\{K_j\}$ and
corresponding cutoff functions $\{\theta_j\}$. However, one can show (see p.~138 of \cite{hw1})
that the *-algebra
generated by the interacting fields does not depend these choices in the sense that
different choices give rise to isomorphic algebras\footnote{These algebras do 
not, of course, define the same subalgebra of $\W^\inv_N[g]$.}.
Furthermore, the local fields and their time ordered products constructed from different 
choices of $\{K_j\}$ and $\{\theta_j\}$ are mapped into each other under this isomorphism. 
In this sense, our algebraic construction of the interacting field theory
is independent of these choices.
Although this is not required in this paper, we remark that 
the above algebras of interacting fields 
can also be equipped with an action of the Poincare group on 
$d$-dimensional Minkowski spacetime by a group of automorphisms transforming the fields in the usual way. 
Thus, we have achieved our algebraic formulation of the interacting quantum field 
theory given by the action~\eqref{60} for an arbitrary, but fixed $N$.

\medskip

We will now take the large $N$ limit of the interacting field theory 
on the algebraic level in a similar way as in the free theory described 
in the previous section, by showing that the (suitably normalized) 
interacting fields can be viewed as formal power series in the 
free parameter $\eps = 1/N$, provided that the `t Hooft coupling 
\ben
\label{71}
g_{\rm t} = gN
\een
is held fixed at the same time. In fact, the suitably normalized  interacting fields will be 
shown to be elements of a subalgebra of the algebra $\X[\eps, g_{\rm t}]$ of 
formal power series in $g_{\rm t}$ with coefficients in $\X[\eps]$. 
 
To begin, we prove a lemma about the dependence upon $N$ of the 
$n$-th order contribution to the the interacting field with cutoff interaction, given by 
the $n$-th retarded product in eq.~\eqref{70}.

\begin{lem}
\label{lemma1}
Let $\eps = 1/N$, $\Phi_i, \Psi_j \in \V^\inv$, let $f_i, h_j \in \D(\mr^d)$, and let $T_i$ respectively
$S_j$ be the number of traces occurring in $\Phi_i$ respectively $\Psi_j$. Then we have 
\ben
\label{72}
R(\prod_{i=1}^n \eps^{-2+T_i+|\Phi_i|/2} f_i \Phi_i; \prod_{j=1}^m \eps^{S_j+|\Psi_j|/2} h_j\Psi_j) = O(1), 
\een
where the notation $O(\eps^k)$ means that the corresponding algebra element of 
$\W^\inv_N$ (supposed to be given for all $N$) can be written as a linear combination of the generators 
$W_{a}(t)$ with $t$ independent of $\eps$, and with 
coefficients of order $\eps^k$. 
\end{lem}
\begin{proof}
For simplicity, we first give a proof of eq.~\eqref{72} in the case $m=1$; the case of general 
$m$ is treated below. Let us define, following~\cite{df}, the ``connected product'' in $\W^\inv_N$ as 
the $k$-times multilinear maps on $\W^\inv_N$ defined recursively by the relation 
\ben
\label{73}
(W_{a_1}(t_1) \cdot \cdots \cdot W_{a_k}(t_k))^{\rm conn} 
\equiv W_{a_1}(t_1) \cdot \cdots \cdot W_{a_k}(t_k)-\sum_{\{1, \dots, k\} = \cup I} \prod^{\rm class}_{I} 
\left( \prod_{j \in I} W_{a_j}(t_j) \right)^{\rm conn},  
\een
where the ``classical product'' $\cdot_{\rm class}$ is the commutative 
associative product on $\W_N^\inv$ defined by 
\ben
\label{74}
W_a(t) \cdot_{\rm class} W_b(s) = W_{ab}(t \otimes s),  
\een
and where the trivial partition $I = \{1, \dots, k\}$ is 
excluded in the sum. We now analyze the $\eps$-dependence of the contracted product, restricting
attention for simplicity first to the case when each of the $W_{a_i}(t_i)$ contains only one trace. 
We use the product formula~\eqref{47} to evaluate the connected product 
$(W_{a_1}(t_1) \cdot \cdots \cdot W_{a_k}(t_k))^{\rm conn}$
as a sum of contributions of the form $\eps^{I} W_{b} (s)$ 
associated with Feynman graphs $\Gamma$, where $I$ is the number of index loops
in the graph, and where $s \in \E_{|b|}'$ does not depend upon $\eps$. 
It is seen, as a consequence of our definiton of the connected product, that 
precisely the connected diagrams occur in the sum. By arguments similar to the one given in 
the previous section, the number $I$ associated
with a given conneceted diagram with $k$ vertices is given by $2-k-H-J$, where $H$ is the 
number of handles of the surface associated with the diagram, and where $J$ is 
the number of traces in the algebraic element $W_{b} (s)$ associated with the 
contribution of that Feynman graph. Consequently, since $H, J \ge 0$, we have 
\ben
\label{75}
(W_{a_1}(t_1) \cdot \cdots \cdot W_{a_k}(t_k))^{\rm conn} = O(\eps^{k-2}) 
\een
when each of the $W_{a_i}(t_i)$ contains only one trace. Now consider the 
retarded product when each of the fields has only one trace, and when the 
supports of the testfunctions $f_i, h$ satisfy
\ben
\label{76}
\supp f_i \cap \supp h = \supp f_i \cap \supp f_j = \emptyset. 
\een
Without loss of generality, we can assume that the supports of 
the $f_i$ have no intersection with either the causal past or 
the causal future of the support of $h$ (otherwise, we write 
each $f_i$ as a sum of two testfunctions with this property). 
Under these assumptions, the retarded product is given by~\cite{df}
\ben
\label{77}
R(\prod_{i=1}^n f_i \Phi_i; h\Psi) = 
\sum_\pi [\Phi_{\pi 1}(f_{\pi 1}), [\Phi_{\pi 2}(f_{\pi 2}), \dots 
[\Phi_{\pi n}(f_{\pi n}), \Psi(h)] \dots ]], 
\een
when the supports of all $f_i$ have no point in common with the causal future
of the support of $h$, and by 0 otherwise. We now use the following 
lemma which we are going to prove below:
\begin{lem}
\label{lemma2}
Let $B, A_1, \dots, A_n \in \W^\inv_N$. Then
\ben
\label{78}
\sum_\pi \left( [A_{\pi n}, [A_{\pi(n-1)}, \dots [A_{\pi 1}, B] \dots ]] \right)^{\rm conn} = 
\sum_\pi [A_{\pi n}, [A_{\pi(n-1)}, \dots [A_{\pi 1}, B] \dots ]].  
\een 
\end{lem}
Since $\eps^{|\Phi_i|/2} \Phi_i(f_i)$ and 
$\eps^{|\Psi|/2} \Psi(h)$ can be written in the
form $W_a(t)$ for some distributions $t$ not depending on $\eps$, it follows by
eqs.~\eqref{75} and~\eqref{77} and the lemma that  
\ben
\label{specialcase}
R(\prod^n_{i=1} \eps^{|\Phi_i|/2} f_i \Phi_i, \eps^{|\Psi|/2} h\Psi) = O(\eps^{n-1})
\een
when the supports of $f_i, h$ satisfy eq.~\eqref{76}, and when each of the fields $\Phi_i, \Psi$ 
contains only one trace. When the testfunctions $f_i, h$ have 
overlapping supports, the formula~\eqref{77} for the retarded products
is not well-defined, or, alternatively speaking, the formula only defines an 
algebra valued distribution on the domain 
\ben
\times^{n+1} \mr^d \setminus \bigcup_{I \subset \{1, \dots, n+1\}} D_I,
\een
where $D_I$ is a ``partial diagonal'' in the product manifold $\times^{n+1} \mr^d$, 
see eq.~\eqref{22}. However, as explained at the end of section~3, we are 
considering a prescription for constructing the time ordered (and hence retarded) 
possessing a Wick expansion of the form eq.~\eqref{Nwickexp} everywhere, including the diagonals. 
The Wick expansion eq.~\eqref{Nwickexp} implies that the $N$-dependence on the diagonals is
identical to that off the diagonals. Equation~\eqref{specialcase} therefore follows immediately 
for all test functions. This proves the desired relation~\eqref{72} 
 when $m=1$ and when all fields contain only one trace.

The situation is a bit more complicated when the fields $\Phi_i, \Psi$ 
contain multiple traces. In that case, we similarly begin by analyzing
the $N$-dependence of the connected product~\eqref{73} when 
the $W_{a_i}(t_i)$ contain multiple traces, so that each $a_i$ now stands 
for a multi index $(a_{i1}, \dots, a_{iT_i})$, where $T_i$ is the number of traces 
in $W_{a_i}(t_i)$, and where $a_{ij}$ is the number of free field factors appearing in 
the $j$-th trace of $W_{a_i}(t_i)$. It is seen that only the following type
of Feynman graphs can occur in the connected product of these algebra elements: 
The valence of the vertices of the graphs are determined by the number of fields $a_{ij}$ appearing
in the $j$-th trace of the $i$-th algebra element. For each fixed $i$, no $a_{ij}$-vertex can be 
connected to a $a_{ik}$-vertex. For fixed $i, l$, there exist indices $j,k$ such the $a_{ij}$-vertex is 
connected to the $a_{lk}$-vertex. Analyzing the $N$-dependence of these graphs arising from 
index contractions along closed index loops in same way as in our analysis of the $N$-dependence
of the algebra product~\eqref{47}, we find that the contributions from 
these Feynman graphs are at most of order 
\ben
\label{80}
O(\eps^{J + 2\sum H_j + \sum V_j - 2C}),  
\een
where $C$ is the number of disconnected components of the surface associated with the graph, 
$J$ is the number of closed index loops containing ``external currents'', $V_j$ is the number of
vertices in the $j$-th disconnected component, and $H_j$ the number of handles (components containing
only a single vertex do not count). Clearly, we have $H_j, J \ge 0$ and we know that
\ben
\label{81}
\sum V_j = \sum T_i -D,  
\een
with $D$ the number of vertices that are not connected to any other vertex.
In order to estimate the number $C$ of connected components of the 
graph, we first assume $D=0$ and imagine the graph obtained by moving all 
the $a_{ij}, j=1, \dots, T_i$ on top of each other for each $i$. The resulting structure
will then only have one connected component, since we know that for fixed $i, l$, 
there exist indices $j,k$ such the $a_{ij}$-vertex is connected to the $a_{lk}$-vertex. 
If we now move the  $a_{ij}, j=1, \dots, T_i$ apart again for a 
given $i$, then it is clear that we will create at most $T_i-1$ new disconnected components. 
Doing this for all $i$, we therefore see that our graph can have at most $1 + (T_1-1) + 
\dots + (T_k-1)$ disconnected components. If $D$ is not zero, then we repeat this argument
for those vertices that are not isolated, and we similarly arrive at the estimate
\ben
\label{82}
C \le 1 - k + \sum T_i -D.  
\een
for the number of disconnected components of any graph appearing in the connected
product~\eqref{73}.
Hence, we find altoghether that 
\ben
\label{83}
(W_{a_1}(t_1) \cdot \cdots \cdot W_{a_k}(t_k))^{\rm conn} = O(\eps^{2k-2-\sum T_i})
\een
when each of the $W_{a_j}(t_j)$ contains $T_j$ traces.
We can now finish the proof in just the same way  
as in the case when all the fields $\Phi_i, \Psi$
contain only a single trace. 

Now let $m$ in eq.~\eqref{72} be arbitrary and consider a situation wherein the supports of 
the testfunctions $f_i, h_j$ satisfy
\ben
\label{84}
\supp f_i \cap \supp h_j = \supp f_i \cap \supp f_j = \supp h_i \cap \supp h_j = \emptyset. 
\een
Without loss of generality, we assume that the support of $f_{i+1}$ has no intersection 
with the causal future of the support of $f_i$. Then it follows from the recursion 
formula~(74) of \cite{df} together with the causal factorization property of the 
time ordered products~\eqref{16} that 
\begin{eqnarray}
\label{85}
R(\prod_{i=1}^n f_i\Phi_i; \prod_{j=1}^m h_j \Psi_j) &=& 
\sum_{\pi} [\Phi_{\pi 1}(f_{\pi 1}), [\Phi_{\pi 2}(f_{\pi 2}), \dots 
[\Phi_{\pi n}(f_{\pi n}), \Psi_1(h_1) \cdots \Psi_m(h_m)] \dots ]] \nonumber \\
&=& \sum_{I_1 \cup \dots \cup I_m = \{1, \dots, n\}} \prod_{k=1}^m \left( \prod_{i \in I_k} 
{\rm ad}(\Phi_i(f_i)) \right)[\Psi_k(h_k)]
\end{eqnarray}
when the supports of all $f_i$ have no point in common with the causal future
of the supports of $h_j$, and by 0 otherwise, and 
where we have set ${\rm ad}(A)[B] = [A, B]$. Since $N^{-|\Phi_i|/2} \Phi_i(f_i)$ and 
$N^{-|\Psi_j|/2} \Psi_j(h_i)$ can be written in the
form $W_a(t)$ for some distribution not depending on $N$, we conclude 
by the same arguments as above that
\ben
\label{86}
\left( \prod_{i \in I_k} 
{\rm ad}(\Phi_i(f_i)) \right)[\Psi_k(h_k)] = 
O(\eps^{2|I_k| - \sum_{i \in I_k} (T_i + |\Phi_i|/2) - (S_k + |\Psi_k|/2)}), 
\een
from which the statement of the theorem follows when the supports of $f_i, h_j$ 
have the properties~\eqref{84}. The general case can be proved from this as above. 

We end the proof of lemma~1 with the demonstration of lemma~\ref{lemma2}: 
Let $A = \sum \lambda_i A_i$ and consider the formal power series expression
\ben
e^{-A} \cdot B \cdot e^A = \sum_{m, n \ge 0} \frac{1}{m! n!} (-A)^m \cdot B \cdot A^n. 
\een
For a fixed $k > 0$, consider the contribution to the sum on the right hand 
side arising from diagrams such that precisely $k$ $A$-vertices are disconnected
from the other $A$- and $B$-vertices. Since disconnected diagrams factorize
with respect to the classical product $\cdot_{\rm class}$ this contribution is 
seen to be equal to 
\ben
\sum_{m, n \ge 0} \frac{1}{m! n!} \sum_{r+s = k} \frac{m!n!}{r!(m-r)!s!(n-s)!} 
((-A)^r \cdot A^s) \cdot_{\rm class} ((-A)^{m-r} \cdot B \cdot A^{n-s}).
\een
But this expression vanishes, due to $\sum_{r+s = k} (-A)^r \cdot A^s/r!s! = 0$, 
showing that $e^{-A} \cdot B \cdot e^A = (e^{-A} \cdot B \cdot e^A)^{\rm conn}$. 
The statement of the lemma is obtained 
by differentiating this expression $n$ times with respect to the parameters $\lambda_i$. 
\end{proof}

Applying the lemma to the retarded products appearing in the definition~\eqref{65} 
of the interacting field with cutoff, 
(i.e., $\Phi_i = \tr \phi^4$, so that $T_i = 1, |\Phi_i| = 4$ in that case), and using 
our assumption $g \propto \eps$ (see eq.~\eqref{71}), we get
\ben
\label{89}
g^n R(\prod^n \theta \tr \phi^4; f\Phi) = O(\eps^{-T-|\Phi|/2}),  
\een
where $T$ is the number of traces in the field $\Phi$. Therefore, since 
the cutoff interacting field is a sum of such terms, we have found 
$\Phi_{\theta V}(f) = O(\eps^{-T-|\Phi|/2})$ for the cutoff interacting fields, 
viewed now as formal power series in the `t Hooft coupling parameter $g_{\rm t}$ rather than $g$.  
For the interacting time ordered products with cutoff, 
we similarly get $T_{\theta V}(\prod \Phi_i(f_i)) = O(\eps^{-\sum T_i + |\Phi_i|/2})$. 
We claim that the same is true for the interacting fields without cutoff:
\begin{prop}
Let $\eps = 1/N$, $\Phi \in \V^\inv$ with $n$ factors of $\phi$ and $T$ traces. Then
\ben
\label{90}
\Phi_{V}(f) = O(\eps^{-T-n/2})
\een
as formal power series in the `t Hooft coupling $g_{\rm t}$. More generally, for the 
interacting time ordered products
\ben
\label{91}
T_{V}(\prod \Phi_i(f_i)) = O(\eps^{-\sum T_i + n_i/2}), 
\een
where $T_i$ is the number of traces in $\Phi_i$, and where 
$n_i$ is the number of factors of $\phi$ in $\Phi_i$. 
\end{prop}
\begin{proof}
According to our definition of the interacting field without cutoff, eq.~\eqref{68}, we must show that 
\ben
\label{92}
\eps^{n/2+T} \cdot U_j \cdot \Phi_{\theta_j V}(f) \cdot U_j{}^{-1} = O(1)
\een
where $\{\theta_j\}$ and $\{U_j\}$ 
are sequences of cutoff functions and unitary elements as in 
our definition of the interacting field, see eq.~\eqref{68}.
We expand $U_j$ and $\Phi_{\theta_j V}(f)$
in terms of the retarded products and use the fact, 
shown in~\cite{df}, that only connected diagrams contribute to each term in the resulting formal 
power series. The $N$-dependence of these terms can then 
be analyzed in a similar fashion as in the proof of lemma~\ref{lemma1} and
gives~\eqref{92}.\footnote{Note, however, that the expansion of $U_j$ itself contains {\em negative} powers 
of $\eps$, i.e., it is {\em not} true that $U_j$ is of $O(1)$
separately.} The proof for the time ordered producs is similar.
\end{proof}

The proposition allows us to view the suitably normalized interacting fields
and their time ordered products as elements of the algebra $\X[\eps, g_{\rm t}]$ of formal power series in 
$g_{\rm t}$ with coefficients in $\X[\eps]$, i.e., we have shown 
\ben
\label{93}
\eps^{T+n/2} \Phi_{V}(f) \in \X[\eps, g_{\rm t}], 
\een
and similarly for the interacting time ordered products\footnote{Note 
that the $\eps$-dependence of the normalization factors necessary to make
the interacting fields and their time ordered 
products elements of $\X[\eps, g_{\rm t}]$ 
differs from that in the free field theory, see~\eqref{58}.}.
We denote by $\B_{V}$ the subalgebra of $\X[\eps, g_{\rm t}]$ generated by the fields~\eqref{93}
and their time ordered products,  
\ben
\label{94}
\B_{V} = \text{ alg$\left\{ \eps^{\sum T_i + |\Phi_i|/2}
\cdot T_V(\prod_i f_i \Phi_i ) 
\,\, \bigg| \,\, f_i \in \D(\mr^d), \Phi_i \in \V^\inv \right\}$} \subset \X[\eps, g_{\rm t}].
\een
By the same arguments as given on p.~138 of~\cite{hw1}, one 
can again prove that, as an abstract algebra, $\B_{V}$ does not depend on the 
choice of the cutoff functions entering in the definition of the interacting field.
Since the algebra $\B_V$ is an algebra of formal power series in $\eps = 1/N$, 
the construction of $\B_{V}$ accomplishes the desired 
algebraic formulation of the $1/N$-expansion for the interacting quantum field theory 
associated with the action~\eqref{60}.

Since the algebra $\B_V$ was constructed perturbatively, it incorporates not only an expansion in $1/N$, but 
also of course a formal  expansion in the coupling parameters. Moreover, one can show that the value of Planck's constant, $\hbar$, 
(set equal to 1 so far) can be incorporated explicitly into the algebra $\A_V$, and it is seen that the classical 
limit, $\hbar \to 0$ can thereby included into our algebraic formulation. Following~\cite{df}, we briefly describe how this is done. 
One first introduces an explicit dependence on $\hbar$ into the algebra product~\eqref{5} in $\W$ by replacing $\Delta_+$ in that product 
formula by $\hbar \Delta_+$. With this replacement understood, $\W$ can now be viewed as a 1-parameter family of 
*-algebras depending on the parameter $\hbar$. It is possible to set $\hbar = 0$ on the algebraic level. In this limit, 
$\W$ becomes a commutative algebra, and $\frac{1}{i \hbar}$ times the commutator defines a Poisson bracket in the limit.
In this way, the 1-parameter family of algebras $\W$ depending on $\hbar$ is seen to be a deformation of the classical Poisson algebra associated
with the free Klein-Gordon field. These consideration can be generalized straightforwardly to the algebras $\X[\eps]$ as well 
as $\X[\eps, g_{\rm t}]$, and we incorporate the dependence on $\hbar$ of these algebras into the new notation 
$\X[\eps, g_{\rm t}, \hbar]$. The algebras of interacting fields, $\B_{V}$, with interaction now taken to be $\frac{1}{\hbar} V$,  
can be seen~\footnote{This is a non-trivial statement, because $\frac{1}{\hbar} V$ contains {\em negative}
powers of $\hbar$. The proof of this statement can be adapted from~\cite{df}.} to be subalgebras of $\X[\eps, g_{\rm t}, \hbar]$, and 
therefore depend likewise on the indicated deformation parameters, 
\ben
\B_{V} = \B_{V}[\eps, g_{\rm t}, \hbar]. 
\een
The interacting field algebras consequently have a classical limit, $\hbar \to 0$, and can thereby be seen 
to be non-commutative deformations of the Poisson algebras of classical (perturbatively defined) 
field observables associated with the action~\eqref{60}, that 
depend on $1/N$ as a free parameter. In this way, the expansion of the large $N$ interacting field theory in terms of 
$\hbar$ is incorporated on the algebraic level, and the classical limit $\hbar \to 0$ can be taken on this level. On the other hand, 
one can show that the vacuum state and the Hilbert space representations of $\B_{V}$ as operators on Hilbert space
{\em cannot} be taken. This demonstrates the strength of the algebraic viewpoint.

\medskip

For a more general interaction 
\ben
\label{95}
V(\phi) = \sum g_i \Phi_i
\een
including interaction 
vertices $\Phi_i \in \V^\inv$ with $T_i$ multiple traces, it follows from lemma~\ref{lemma1} that 
the interacting field will still satisfy eq.~\eqref{93}, provided that the coupling constants $g_i$ tend to zero
for large $N$ in such a way that the corresponding `t Hooft parameters $g_{i \rm t}$ defined by
\ben
\label{96}
g_{i \rm t} = g_i N^{T_i + |\Phi_i|/2 -2}
\een
remain fixed (note that \eqref{71} is the special case $\Phi_i = \tr \phi^4$ of this relation). 
Thus, if the coupling constants $g_i$ are tuned in the prescribed way, 
the  interacting field algebra $\B_{V}$ is defined as a subalgebra of the 
algebra $\X[\eps, g_{1 \rm t}, g_{2 \rm t}, \dots]$ 
of formal power series in the `t Hooft coupling parameters
with coefficients in $\X[\eps]$.

\medskip

The perturbative expansion of the interacting fields~\eqref{93}
defined by the interaction~\eqref{61} as an element of $\B_{V}$ is organized
in terms of Feynman graphs that are associated with Riemannian surfaces, where 
contributions from genus $H$ surfaces are suppressed by a factor $\eps^H$. 
To illustrate this in an example, consider the interacting field 
$\eps^{3/2} (\tr \phi)_{\theta V}$ with cutoff interaction $\theta(x) V$. 
In order to have a compact notation for the decomposition of 
the $n$-th order retarded product occuring in the 
perturbative expansion of this interacting field into contributions
associated with Feynman graphs, we first consider a corresponding 
retarded product occurring in $\phi_{\theta V}$ in the theory of a single scalar field with 
interaction $\theta(x) V$, where $V = g\phi^4$. Such a retarded product can be decomposed in the form\cite{s}
\ben
\label{97}
R(V(y_1) \cdots V(y_n); \phi(x)) = g^n
\sum_{{\rm graphs} \,\, \Gamma} r_\Gamma(y_1, \dots, y_n; x) :\phi^{a_1}(y_1) \cdots \phi^{a_n}(y_n):_H.
\een
The sum is over all connected graphs $\Gamma$ with 4-valent vertices $y_i$ and a 1-valent 
vertex $x$, and $a_i$ is the number of external legs (i.e., lines with open ends) attached to the vertex $y_i$. 
The $r_\Gamma$ are c-number distributions associated with the graph which are determined by 
appropriate Feynman rules.

We now look at  a correpsonding retarded product occuring in the perturbative expansion of the 
corresponding field $\eps^{3/2}(\tr\phi)_{\theta V}$
in the large $N$ interacting quantum field theory with $V = g \tr \phi^4$.  By an analysis analogous
to the one given in the proof of lemma~1, it 
can be shown that such a retarded product can be written as a sum of 
contributions from individual Feynman graphs as follows:
\begin{multline}
\label{98}
\eps^{3/2} R(V(y_1) \cdots V(y_n); \tr \phi(x)) = 
g_{\rm t}^n \sum_{{\rm genera}\,\,H} \eps^H \sum_{{\rm graphs} \,\, \Gamma} 
\eps^{f/2 + T} 
\\
\cdot r_{\Gamma}(y_1, \dots, y_n; x) 
\,\,: \tr \prod_i \phi(y_i)               
\cdots \tr \prod_j \phi(y_j):_H. 
\end{multline}
The expression on the right side is to be understood as follows: 
$g_{\rm t}$ is the `t Hooft coupling~\eqref{71}. 
The sum is over all distinct Feynman graphs $\Gamma$ that occur in the corresponding expansion~\eqref{97}
in the theory with only a {\em single} scalar field, and the c-number distributions
$r_\Gamma$ are identical to the ones appearing in that expansion. The sum over graphs is 
subdivided into contributions grouped together according to their topology specified by the 
genus, $H$, of the graph, defined as the number of handles of the surface $\S$ obtained
by attaching faces to the closed index loops occuring in the given graph 
(we assume that a double line notation as described in section~3 is used for the propagators 
and the vertices). The external legs are incorporated by capping
off each such external line connected to $y_k$ and ending on the index pair $ij'$ with an 
``external current'' $\phi_{ij'}(y_k)$. The external currents are collected in the normal ordered
term appearing in eq.~\eqref{98}, where each trace corresponds to following through the 
index line to which the currents within that trace belong. The number of traces in 
such a normal ordered term is denoted $T$, and the number of factors of $\phi$ is 
denoted $f$.

The same remarks also apply to the perturbative expansion of the more general gauge invariant fields 
$\eps^{|\Phi|/2 + T}\Phi_{\theta V}$ in the large $N$ theory. A similar expansion is also valid 
for the corresponding fields without cutoff $\theta$. 
Moreover, $V$ may be replaced by an arbitrary (possibly non-renormalizable)
local interaction of the form~\eqref{95}, provided that the couplings are tuned in the large $N$ limit
in the manner prescribed in eq.~\eqref{96}.

\section{Renormalization group}

Our construction of the interacting field theory given in the previous section 
is equally valid for interactions $V$ that are renormalizable by 
the usual power counting criterion as well as for non-renormalizable theories. 
Let us sketch
how the distinction between renormalizable and non-renormalizable theories appears in 
the algebraic framework that we are working in. For simplicity, let us first consider the 
theory of a single hermitian scalar field, $\phi$. We take the action of this scalar field to consist
of a free part given by eq.~\eqref{1}, and an interaction given by 
$V = \sum g_i \Phi_i$, which might be renormalizable or non-renormalizable.
(As above, $\Phi_i$ are monomials in $\phi$ and its derivatives.)
The difference between renormalizable $V$ and non-renormalizable $V$ 
shows up in the perturbatively defined interacting quantum field theory as follows: Our definition of 
interacting fields depends on a prescription for defining the Wick powers and their time ordered products
in the free theory, which is given by a map $T$ with the properties (t1)--(t8) specified in section~3. 
As explained there, these properties do not, in general, determine the time ordered products
(i.e., the map $T$) uniquely, and this consequently
leaves a corresponding ambiguity in the definition of the interacting fields. However, as first shown\footnote{
The constructions in~\cite{hw1} were actually given in the more general context of an interacting (scalar) 
field theory on an arbitrary globally hyperbolic curved spacetime. An explicit treatment 
of the special case of Minkowski spacetime was recently given in~\cite{klmi}.} in~\cite{hw1}, 
the algebra of interacting fields associated with interaction $V$ constructed from a given prescription $T$ is 
isomorphic to the algebra constructed from any other prescription, $T'$, provided the interaction is also changed
from $V$ to $V' = \sum g'_i \Phi_i$, where each of the modified couplings $g'_i$ is a 
suitable formal power series in the 
couplings $g_1, g_2, \dots$. Renormalizable theories are 
characterized by the fact that $V'$ always has the same form as $V$, modulo terms of the form 
already present in the free Lagrangian. 

If $\Phi_V$ are the interacting fields constructed from the 
interaction $V$ using the first prescription for defining time ordered products in the free theory, 
and if $\Phi'_{V'}$ are the fields constructed from the interaction 
$V'$ and the second prescription, then the above isomorphism, let us call it $R$, can be 
shown~\cite{hw1} to be of the form
\ben
\label{99}
R: \Phi_{i V}^{} \to \sum_j Z_{ij} \cdot \Phi'_{j V'}, 
\een
where we have omitted the smearing functions for simplicity. The ``field strength renormalization''
constants $Z_{ij}$ are formal power series in $g_1, g_2, \dots$. For renormalizable theories, one
can show that there will appear only
finitely many terms in the sum on the right side. The possible terms are restricted in that 
case by the requirement that the fields $\Phi_j$ on the right side cannot have a greater 
engineering dimension than the field $\Phi_i$ on the left side. 
In a non-renormalizable 
theory, no such restriction occurs. The map $R$ together with the transformation $V \to V'$ 
corresponds to the ``renormalization group'' in other approaches. Since the interactions $V$ might be viewed
as elements of the abstract vector space $\V$ spanned by the field monomials $\Phi$, 
we may view the renormaliztion group as providing 
a map $\V \to \V$. The subspace of renormalizable interaction vertices $V \in \V$ thus correpsonds precisely
to the largest finite dimensional subspace of $\V$ that is invariant under all renormalization 
group transformations.

One can in particular consider the special case in which the alternate prescription $T'$ is related to 
the original prescription, $T$, for defining the time ordered products in the free theory 
by a multiplicative change of scale (with multiplication factor $\lambda > 0$), i.e., $T'$ is given 
terms of $T$ by eq.~\eqref{tlambda}. In that case, 
we obtain a family of isomorphisms $R(\lambda)$ labelled by the parameter $\lambda$, 
together with one-parameter families $g_i' = g_i(\lambda)$, $V' 
= \sum g_i(\lambda) \Phi_i$ and $Z_{ij}(\lambda)$ (for details, we refer to \cite{hw1}). 
By the almost homogeneous scaling behavior of the time ordered products in the free theory, 
eq.~\eqref{tlog}, it the follows that each term appearing in the power series expansions of 
$g_i(\lambda)$ and $Z_{ij}(\lambda)$ depends at most {\em polynomially on $\ln \lambda$}. 
The functions $\lambda \to g_i(g_1, g_2, \dots, \lambda)$ define 
the ``renormalization group flow'' of the theory, which may be viewed as a 1-parameter family
(in fact, group) of diffeomorphisms on $\V$. Thus, our formulation of the 
renormalization group flow is that a given way of defining the interacting fields $\Phi_V$ (i.e., 
using a given renormalization prescription) is equivalent, via the isomorphism $R(\lambda)$, to 
defining the fields $\Phi_{V'}'$ via the ``rescaled'' prescription --- denoted by ``prime'' --- 
obtained from the previous prescription 
by changing the ``scale'' according to eq.~\eqref{tlambda}, provided that the 
interaction is at the same time modified to $V' = \sum g_i(\lambda) \Phi_i$.

We can re-express this renormalization group flow in a somewhat more transparent way 
by noting that, from eq.~\eqref{tlambda}, the rescaled prescription (i.e., the ``primed'' prescription
appearing in the renormalization group flow~\eqref{99}) 
is given in terms of the original one (up to the isomorphism $\sigma_\lambda$)
simply by appropriately rescaling the mass, the field strenght
and the coordinates in the time ordered products in the free theory. Thus, by composing $R(\lambda)$ with $\sigma_\lambda$, 
we get the following equivalent version of our algebraic formulation of the renormalization group flow:
Let $\A_V(U)$ be the algebra of interacting fields smeared with testfunctions supported in a region 
$U \subset \mr^d$ of Minkowski space. Then $\rho_\lambda = R(\lambda) \circ \sigma_\lambda$ is given by
\begin{eqnarray}
\label{renormalize'}
\rho_\lambda: \A_V^{(m)}(\lambda U) &\to& \A^{(\lambda^{-1} m)}_{V(\lambda)} (U), \nonumber \\
\Phi_{i V}(\lambda x) &\to& \sum_j \lambda^{-d_j} Z_{ij} (\lambda) \cdot \Phi_{j V(\lambda)}(x)
\end{eqnarray}
and is again an isomorphism, where we are now indicating the dependence of the algebras upon the mass parameter, $m$. 
Here, $V(\lambda) = \sum \lambda^{-\delta_i} g_i(\lambda) \Phi_i$, 
where $d_i$ is the engineering dimension of the field $\Phi_i$ and $\delta_i$ the engineering
dimension of the corresponding coupling $g_i$, 
and the functions $Z_{ij}(\lambda), g_i(\lambda)$ are as in eq.~\eqref{99}.
Stated differently, the action of $\rho_\lambda$ is 
described as follows: If the argument of an interacting field is  rescaled by $\lambda$, this is equivalent
via $\rho_\lambda$ to a redefinition of the interaction, $V \to V(\lambda)$ together with a suitable 
redefinition of the field strength by the matrix $\lambda^{-d_j} Z_{ij}(\lambda)$. We also note explicitly that 
eq.~\eqref{renormalize'} makes reference to only {\em one} given renormalization prescription.

\medskip
 
An important feature of our algebraic formulation of the renormalization group flow  
is that it is given directly in terms of the interacting field {\em operators} which are members of the algebra $\A_V$, rather than 
in terms of the correlation functions of these objects, as is normally done. Of course, one can always 
apply a state (i.e., a normalized linear functional on the field algebra) to the relation~\eqref{99} and 
thereby obtain a relation for the behavior of the Green's functions under a rescaling. Our algebraic formulation
makes it clear that {\em the 
existence of the renormalization group flow is an algebraic property of the theory, i.e. it is encoded in 
the local algebraic relations between the quantum fields. It has nothing to do a priori with the vacuum state or e.g. the
superselection sector of the theory.}
Besides 
offering a conceptually new perspective on the nature of the renormalization group flow, 
our algebraic formulation has the advantage that, since the construction is essentially of a local nature, 
it works regardless of what the infra-red behavior of the theory is. This makes the algebraic approach 
superior e.g. in curved spacetime~\cite{hw1}, where there is no preferred vacuum state, and where 
moreover the infra-red behavior of generic states is very difficult to control (and at any rate, 
depends upon the behavior of the spacetime metric at large distances).

\medskip

The statements just made for the theory of a single, scalar field carry over straightforwardly
to a multiplet of scalar fields. In particular, they are true for  
the theory of a field $\phi$ in the ${\bf N} \otimes \bar {\bf N}$ representation of the group $U(N)$ 
with action~\eqref{60}, for any arbitrary but fixed $N$. 

The aim of the present section is to show that, for gauge invariant interactions, the algebraic 
formulation of renormalization group carries over in a meaningful way in the limit of large $N$, 
or more properly, that the renormalization group it can be defined in the 
sense of power series in $\eps = 1/N$, with positive powers. For this, 
consider two different prescriptions $T$ and $T'$ for defining Wick powers and time 
ordered products in the free theory satisfying (t1)--(t8), as well as eq.~\eqref{59a}.
An explicit construction of such a prescription was given at the end of section~3, but we 
will not need to know the details of that construction here. 
For a given gauge invariant interaction 
$V = \sum g_i \Phi_i \in \V^\inv$ (renormalizable or non-renormalizable), let 
$\B_V$ respectively $\B'_V$ be the algebras of interacting field observables 
constructed via the two prescriptions, each of which is a subalgebra of $\X[\eps, g_{1 \rm t}, 
g_{2 \rm t}, \dots]$, where $g_{i \rm t}$ are the `t Hooft coupling parameters related
to the couplings $g_i$ in the interaction via formula~\eqref{96}. Let the interacting quantum fields in these algebras
be $\eps^{|\Phi|/2 + T} \Phi_V$, respectively $\eps^{|\Phi|/2 + T} \Phi_V'$ 
($T$ the number of traces), defined as formal power
series in $\eps$ and the `t Hooft parameters $g_{i \rm t}$.  

\begin{prop}
For any given $V = \sum g_i \Phi_i \in \V^\inv$ there exists a 
$V' = \sum g_i' \Phi_i \in \V^\inv$ and a *-isomorphism 
\ben
\label{100}
R: \B_V \to \B_{V'}'
\een
such that $g_i' = g_{i \rm t}' \eps^{T_i + |\Phi_i|/2 -2}$ ($T_i$ is the number of 
traces in the field $\Phi_i$), with 
\ben
\label{101}
g_{i \rm t}' = g_{i \rm t}'(g_{1 \rm t}, g_{2 \rm t}, \dots, \eps)
\een 
a formal power series in $g_{i \rm t}$ and $\eps$ (i.e., containing only positive powers of $\eps$). 
The action of $R$ on a local field is given by 
\ben
\label{102}
R: \underbrace{\eps^{|\Phi_i|/2 + T_i}\Phi_{i V}^{}}_{\in \B_V} 
\to \sum_j \Z_{ij} \cdot \underbrace{\eps^{|\Phi_j|/2 + T_j} \Phi'_{j V'}}_{\in \B_{V'}'}, 
\een
where $\Z_{ij}$ are formal power series in $g_{1 \rm t}, g_{2 \rm t}, \dots$ and $\eps$, and 
where $T_i$ is the number of traces in $\Phi_i$. [Recall that the $\eps$-normalization factor in 
expressions like $\eps^{|\Phi|/2 + T} \Phi_V$ in the above equation is precisely the factor needed to 
make the latter an element of $\A_V$.] A similar
formula holds for the time ordered products. 

Moreover, if the ``prime'' prescription is related 
to the ``unprime'' prescription via a multiplicative change of scale (with multiplication factor 
$\lambda$), then each term in the expansion of $g_{i \rm t}'(\lambda)$ and 
$\Z_{ij}(\lambda)$ depends at most polynomially on $\ln \lambda$, e.g.,  
\ben
\Z_{ij}(\lambda) = \sum_{a_1, a_2, \dots, h \ge 0} 
z_{ij, a_1 a_2 \dots  h}(\ln \lambda) g_{1 \rm t}^{a_1} g_{2 \rm t}^{a_2} \dots \eps^h
\een
where the $z_{ij, a_1 a_2 \dots h}$ are polynomials in $\ln \lambda$. 
\end{prop}
\begin{proof}
For any given, but fixed $N$, one can show by the same arguments as in~\cite{hw2} 
that any two prescriptions $T$ and $T'$ for 
defining time ordered products with properties (t1)--(t8) are related to each other in the following way:
\ben
\label{103}
T'\left(\prod_{i=1}^n f_i \Phi_i\right) = T\left(\prod_{i=1}^n f_i \Phi_i\right) + 
\sum_{\cup_j I_j = \{1, \dots, n\}} T\left(\prod_{j} \delta_{|I_j|} (\prod_{k \in I_j} f_k \Phi_k) \right). 
\een
Here, the following notation has been introduced: The sum runs over all partitions of the set 
$\{1, \dots, n\}$, excluding the trivial partition. The $\delta_k$ are maps 
\ben
\label{104}
\delta_k : \otimes^k \D(\mr^d; \V^\inv) \to \D(\mr^d; \V^\inv), 
\een
characterizing the difference between $T$ and $T'$ at order $k$. 
The maps $\delta_k$ have the form\footnote{Note that $\delta_1$ is not 
the identity, since we are allowing ambiguities in the definition of
Wick powers, rather than defining them by normal ordering.}
\ben
\label{105}
\delta_k(\prod_{i=1}^k f_i \Phi_i) = \sum_i F_{k, i} \Psi_i, 
\een
where the functions $F_{k,i}$ are of the form
\ben
\label{106}
F_{k, i}(x) = \sum_{(\mu_1) \dots (\mu_k)} c_{k, i}{}^{(\mu_1) \dots (\mu_k)} \prod_k \partial_{(\mu_k)} f_k(x),  
\een
with each $(\mu_i)$ denoting a symmetrized spacetime multi index $(\mu_{i1} \dots \mu_{is})$, 
and with each $c_{k, i}{}^{(\mu_1) \dots (\mu_k)}$ denoting a Lorentz invariant tensor field (independent 
of $x$). Let us define a $V'$ in $\V^\inv[g_1, g_2, \dots]$ (the space of formal power series in $g_i$ 
with coefficients in $\V^\inv$) by 
\ben
\label{107}
V' = \lim_{j \to \infty} \sum_{k \ge 1} \frac{i^k}{k!} \delta_k(\prod^k \theta_j V), 
\een
where $\{\theta_j\}$ represents any series of cutoff functions that are equal to 1 in compact sets $K_j$ exhausting
$\mr^d$ in the limit as $j$ goes to infinity. 
Then, for any given but fixed $N$, the result~\cite{hw2} establishes the existence of an 
isomorphism $R$ between the algebras of interacting field observables associated 
with the two prescriptions satisfying eq.~\eqref{99} for some set of 
formal power series $Z_{ij}$ in $g_1, g_2, \dots$, where the fields in that equation 
are now given by gauge invariant expressions in $\V^\inv$.

In order to prove the theorem, we must show that the interaction $V'$ and the factors $Z_{ij}$ appearing in 
the automorphism $R$ have the $N$-dependence specified by eqs.~\eqref{99}
respectively~\eqref{98}. 
This will guarantee that the above automorphisms $R$ defined separately for each $N$ given rise 
to a corresponding automorphism of the interacting field algebras, viewed now as depending on $\eps = 1/N$
as a free parameter. 
 
In order to analyze the $N$-dependence of $V'$, let us 
consider the prescription $T''$ defined by $T'$ when applied to $k$ factors or more, and defined by 
eq.~\eqref{103} when applied to $n \le k-1$ factors. Then, by definition, the prescriptions $T$ and $T''$ 
will agree on $n \le k-1$ factors, and 
\ben
\label{108}
T''\left(\prod_{i=1}^k f_i \Phi_i \right) - T\left(\prod_{i=1}^k f_i \Phi_i\right) = \sum_i \Psi_i(F_{k, i}), 
\een
where $F_{k, i}$ is as in eq.~\eqref{106}. If $S_i$ is the number of traces in the field $\Psi_i$, then 
we claim that 
\ben
\label{109}
\Psi_i(F_{k, i}) = O(1/N^{S_i+2k-2-\sum T_j+|\Phi_j|/2}),  
\een
where we recall that an algebra element $A \in \W_N^\inv$ given for all $N$ is said to be $O(1/N^h)$ if it can 
be written as $1/N^h$ times a sum of terms of the form $W_a(t_a)$, with each $t_a \in \E'_{|a|}$ 
depending only on positive powers of $1/N$. Using eq.~\eqref{58}, eq.~\eqref{109} is equivalent to
\ben
\label{110}
F_{k, i} = O(1/N^{S_i+|\Psi_i|/2+2k-2-\sum T_j+|\Phi_j|/2}).
\een 
Assuming that this has been shown, we get the statement~\eqref{98}
about the $N$-dependence of $V'$ by plugging this relation into eqs.~\eqref{104}, \eqref{106} and~\eqref{107}, 
and using the definition of 
the `t Hooft couplings, eq.~\eqref{96}. In order to show~\eqref{110}, let us begin by introducing the 
``connected time ordered product'' as the map $T^{\rm conn}: \otimes^k \D(\mr^d, \V^\inv) \to 
\W^\inv_N$ defined recursively in terms of $T$ by 
\ben
\label{111}
T^{\rm conn}\left(\prod_{i=1}^n f_i \Phi_i \right)  
\equiv  T\left(\prod_{i=1}^n f_i \Phi_i \right)-\sum_{\{1, \dots, n\} = \cup I} \prod^{\rm class}_{I} 
T^{\rm conn}\left( \prod_{j \in I} f_j \Phi_j \right),  
\een
where the ``classical product'' $\cdot_{\rm class}$ is the commutative 
associative product on $\W_N^\inv$ defined by eq.~\eqref{74}, 
and where the trivial partition $I = \{1, \dots, k\}$ is 
excluded in the sum. By definition, we have $T^{\prime\prime \rm conn} = T^{\rm conn}$ when 
acting on $n \le k-1$ factors, because $T'' = T$ in that case. This implies that 
we can alternatively write $\sum_i \Psi_i(F_{k, i})$ in eq.~\eqref{108} as the corresponding difference
of connected time ordered products. By a line of arguments similar to the proof 
of eq.~\eqref{83} in lemma~\ref{lemma1} using that only connected Feynman diagrams 
contribute to the connected time ordered products, it can be seen that
have 
\ben
\label{112}
T^{\rm conn} \left(\prod_{i=1}^k f_i \Phi_i \right) = 
\sum_j (1/N)^{j+2k-2-\sum T_l+|\Phi_l|/2} \sum_{a = (a_1, \dots, a_j)} W_a(t_a(\otimes_i f_i)),  
\een
where each $t_a$ is a linear map 
\ben
\label{113}
t_a: \otimes^k \D(\mr^d) \to \E'_{|a|}
\een
which can contain only positive powers of $1/N$. 
A completely analogous estimate holds for $T^{\prime\prime \rm conn}$, with $t_a$
replaced by maps $t_a''$ with the same property. By eq.~\eqref{108} (with the time 
ordered products replaced by the connected products in that equation), we therefore
find
\ben
\label{114}
\sum_i \Psi_i(F_{k, i}) = 
\sum_j (1/N)^{j+2k-2-\sum T_l+|\Phi_l|/2} \sum_{a = (a_1, \dots, a_j)} W_a(s_a), 
\een
where we have set $s_a = t_a - t_a''$. We now write the expressions appearing 
on the left side as $\Psi_i(F_{k, i}) = W_a(u_a)$, where the distributions $u_a$ 
are related to $\Psi_i$ and $F_{k, i}$ via a relation of the form~\eqref{19} and~\eqref{20}. If we 
now match the terms on both sides of this equation and use the linear independence 
of the $W_a$'s, we obtain the desired estimate~\eqref{109}. As already explained, this proves the 
desired $N$-dependence of $V'$. 

The proof that the field strength renormalization factors $Z_{ij}$ in eq.~\eqref{99}
have the desired $N$-dependence expressed in eq.~\eqref{102} is very similar to the 
proof that we have just given, so we only sketch the argument. For a given, but fixed $N$, 
the factors $Z_{ij}$ are defined implicitly by the relation
\ben
\label{115}
\lim_{l \to \infty} \sum_{k \ge 0} 
\frac{i^k}{k!}\delta_{k+1}(f\Phi_i, \prod^k \theta_l V) = \sum_j Z_{ij} f\Phi_j. 
\een
The desired $N$-dependence of the field strength renormalization factors implicit in eq.~\eqref{102}
is equivalent to 
\ben
\label{116}
Z_{ij}(\eps, g_1, g_2, \dots) = \eps^{|\Phi_i|/2 + T_i - |\Phi_j|/2 - T_j} \cdot \Z_{ij}(\eps, g_{1 \rm t}, 
g_{2 \rm t}, \dots), 
\een
where $\Z_{ij}$ are formal power series in the `t Hooft parameters and $\eps$ (i.e., depending only 
on positive powers of $\eps$), and where $T_i$ is the number of traces in the field $\Phi_i$. 
In order to prove this equation from the definition~\eqref{115}, one
proceeds by analyzing the $N$-dependence of the maps $\delta_{k+1}$ in the same way as above. 

The desired {\em polynomial} dependence of the coefficients of $\Z_{ij}(\lambda)$ and $g_{i \rm t}'(\lambda)$
on $\ln \lambda$ when $T'$ arises from $T$ via a scale transformation follows 
as in~\cite{hw1} from the almost homogeneous scaling behavior~\eqref{tlog} of the time ordered prodcuts in 
the free theory. 
\end{proof}

\section{Reduced symmetry}

In the previous sections, we have constructed interacting field algebras associated 
with a $U(N)$-invariant action as a power series in $1/N$. 
Instead of considering $U(N)$-invariant actions, 
one can also consider actions that are only invariant under 
some subgroup. In the present section we will consider actions of the form eq.~\eqref{64} 
in which the free part of the action is invariant under the full $U(N)$-group, 
and in which the interaction term $V$ is now invariant only under a subgroup $G$ of the form 
\ben
\label{117}
G = U(N_1) \times \cdots \times U(N_k) \subset U(N), 
\een
where $\sum N_\alpha = N$. Since the perturbative construction of a quantum field theory 
with such an interaction involves Wick powers and time ordered products in the free theory
that are not invariant under the full $U(N)$ symmetry group but only under the subgroup, 
we begin by describing the algebra 
\ben
\label{118}
\W^G_N = \{ A \in \W_N \mid \alpha_U(A) = A, \quad \forall U \in G\}
\een
of observables invariant under $G$ of which these fields are elements. 
If we define $P_\alpha$ to be the projection matrix corresponding to the $\alpha$-th factor in the 
product~\eqref{117}, then it is easy to see that $\W^G_N$ is spanned by expressions of the form
\begin{multline}
\label{119}
W_{a, \alpha}(t) = \\
\frac{1}{N^{|a|/2}} \int
: \tr \left( \prod_{i_1 \in I_1} \phi(x_{i_1}) P_{\alpha_{i_1}} \right)
\cdots 
\tr \left( \prod_{i_T \in I_T} \phi(x_{i_T}) P_{\alpha_{i_T}} \right):
t(x_1, \dots, x_{|a|}) \prod_j d^d x_j .  
\end{multline}
Here, $a$ represents a multi index $(a_1, \dots, a_T)$, $\alpha$ represents 
a multi index $(\alpha_1, \dots, \alpha_{|a|})$, the $I_j$'s are mutually disjoint
index sets with $a_j$ elements 
each such that $\cup_j I_j = \{1, \dots, |a|\}$, and $t$ is a distribution in the space $\E'_{|a|}$.

The large $N$ limit of the algebras 
$\W_N^G$ can be taken in a similar way as in the case of 
full symmetry described in section~3, provided that the ratios  
\ben
\label{120}
s_\alpha = N_\alpha/N 
\een
have a limit. As above, the large $N$ limit is incorporated in the construction of a suitable
algebra $\X_{\s}[\eps]$, depending now on the ratios $\s = (s_1, \dots, s_k)$, 
of polynomial espressions in $\eps$ whose coefficients are 
given by generators $W_{a, \alpha}(t)$. To work out the algebra
product between two such generators as a power series in $\eps = 1/N$, 
it is useful again to consider first the simplest case $d=0$ corresponding to the  
matrix model given by the action functional~\eqref{31}, and by considering the 
product of the matrix generators
\ben
\label{121}
W_{a, \alpha} = 
\frac{1}{N^{|a|/2}} \, : \prod^T_i \tr \bigg( \underbrace{M P_{\alpha_i} M P_{\alpha_{i+1}}
\cdots M P_{\alpha_k}}_{\text{$a_i$ factors of $M$}} \Bigg) :, 
\een
corresponding to the algebra elements~\eqref{119}. We expand the 
product of two such generators in terms of Feynman graphs as in section~3, 
the only difference being that the vertices corresponding to the traces in eq.~\eqref{121}
now also contain projection operators $P_\alpha$. We take this into account by modifying 
our notation of these vertices by indicating also the projectors adjacent to the vertex. 
As an example, consider the generator with a single trace given by 
\ben
\label{example}
W_{3, (\alpha_1, \alpha_2, \alpha_3)} = \frac{1}{N^{3/2}} : \tr P_{\alpha_1} M  P_{\alpha_2} M P_{\alpha_3} M :.
\een
This generator will contribute a 3-valent vertex drawn in the following picture:

\vspace{1cm}
\begin{center}
\begin{fmffile}{graph5}
\begin{fmfgraph*}(100,80) 
\fmfpen{thick}
\fmfleft{i1,i2}
\fmfright{o1}
\fmf{dbl_plain,label=$P_{\alpha_1}$,label.side=left}{i1,w1}
\fmf{dbl_plain,label=$P_{\alpha_2}$,label.side=left}{i2,w1}
\fmf{dbl_plain,label=$P_{\alpha_3}$,label.side=left}{o1,w1}
\fmflabel{$ij'$}{i1}
\fmflabel{$jk'$}{i2}
\fmflabel{$ki'$}{o1}
\end{fmfgraph*}
\end{fmffile}
\end{center}

A closed loop of index contractions
occurring in a diagram associated with the product
$W_{a, \alpha} \cdot W_{b, \beta}$ will contribute a factor of 
\ben
\tr \left( \prod_{\gamma \in \{\alpha_i, \beta_j\}} P_\gamma \right),  
\een
where the product is over all projectors that are encountered when following the index loop. But the projection
matrices $P_\gamma$ are mutually orthogonal, $P_{\gamma'} P_\gamma = \delta_{\gamma\gamma'} P_\gamma$, 
so this factor is given by $N_\gamma$
if the projectors in the index loop are all equal to some $\gamma$, and vanishes otherwise. If we 
let $I_\alpha$ be the number of index loops in a given graph containing only projections
on the $N_\alpha$-subspace, then we consequently get 
 \ben
\label{123}
W_{a, \alpha} \cdot W_{b, \beta} 
= \sum_{\rm graphs} s_1^{I_1} \dots s_k^{I_k} 
\eps^{J + \sum H_k + \sum(V_k-2)}
\prod_{\text{lines $(k,l)$}} \frac{1}{m^2} 
\cdot W_{c, \gamma}, 
\een
where the sum is only over graphs whose index loops contain only one kind of projectors, 
and where $\eps = 1/N$ as usual. 
As in the previous section, $W_{\gamma, c}$ arises from the graphs with loops containing $c_j$ external
currents each, and $c$ denotes the multi index $(c_1, c_2, \dots)$. The new feature is that
each of these loops now also contains projection operators $P_\gamma$. 

As in the previous section, we can generalize the considerations leading to formula~\eqref{123} to determine the 
product of algebra elements $W_{a, \alpha}(t), t \in \E'_{|a|}$ when the spacetime dimension is not zero. 
We thereby obtain a family of algebras $\X_\s[\eps]$ depending analytically 
on the ratios $\s = (s_1, \dots, s_k)$. Since by definition $0 \le s_\alpha \le 1$ and 
$\sum s_\alpha = 1$, the tuples $\s$ can naturally be viewed as elements 
of the standard $(k-1)$-dimensional simplex 
\ben
\label{124}
\Delta_{k-1} = \{(s_1, \dots, s_k) \in \mr^k \mid 0 \le s_\alpha \le 1, \sum s_\alpha = 1\}. 
\een
Thus, our construction of the algebras associated with the reduced symmetry 
group yields a bundle
\ben
\label{125}
\sigma_{k-1} : \Delta_{k-1} \to {\rm Alg}, \quad \s \to \X_\s[\eps]
\een
of algebras with every point of the standard $(k-1)$-symplex for every $k$. The 
parameters $\s$ interpolate continuously between situations of different symmetry. 
If $\Delta_l$ is a face of $\Delta_k$, (so that $l < k$), then the assignment 
fulfills the ``self-similar'' restriction property
\ben
\label{126}
\sigma_k \restriction \Delta_l = \sigma_l. 
\een
The extremal points of $\Delta_k$ (i.e., the zero-dimensional faces) correspond
to full symmetry, i.e., the restriction of $\sigma_k$ to these points yields 
the algebras $\X[\eps]$ constructed in section~3. 

\medskip

We now repeat the construction of the interacting field algebras $\B_V$, 
$V = \sum g_i \Phi_i$, with each $\Phi_i$  an expression in the field that is 
invariant under the reduced symmetry group. We denote the vector 
space of such formal expressions by  
\ben
\label{127}
\V^\inv_k = \text{span} 
\left\{ \Phi = \prod_i \tr \left( \prod^{a_i} (P_{\alpha_l} \partial_{\mu_1} \cdots \partial_{\mu_j} 
\phi) \right), \alpha_l = 1, \dots, k \right\}, 
\een
(note that $\V^\inv_1$ can naturally be identified with 
$\V^\inv$ in the notation introduced earlier). 
For an arbitrary but fixed $N$, a gauge invariant interaction $V = \sum g_i \Phi_i \in \V^\inv_k$ 
gives rise to corresponding interacting quantum fields $\Phi_V$ and their
time ordered products as elements in the corresponding algebra $\W^G_N[g_1, g_2, \dots]$. 
The limit $N \to \infty$ can be taken on the algebraic level in the same way
as in the case of full symmetry described in section~4, provided that the ratios $s_\alpha = N_\alpha/N$ are held fixed, 
and provided that the couplings are tuned as in~\eqref{96}. This construction 
directly leads to an algebra $\B_{V, \s}$  
of formal power series in $\eps = 1/N$ as well as the  
`t Hooft couplings $g_{i \rm t}$ of which the smeared normalized gauge invariant
interacting fields $\eps^{|\Phi|/2 + T} \Phi_V(f)$ and their time ordered products 
are elements. The algebra $\B_{V, \s}$ is 
now a subalgebra of the algebra $\X_\s[\eps, g_{1{\rm t}}, g_{2{\rm t}}, \dots]$ of 
formal power series  in the `t Hooft couplings, with coefficients in the  
algebra $\X_\s[\eps]$. 

We have constructed in this way a family algebra $\B_{V, \s}$ parametrized by 
deformation parameters $\s$, i.e., a bundle
\ben
\label{128}
\sigma_{V, k}: \Delta_{k-1} \to {\rm Alg}, \quad \s \to \B_{V, \s}, 
\een
where $\Delta_{k-1}$ is the $(k-1)$-dimensional standard simplex~\eqref{124} of which the $\s$ are elements. 
These deformation parameters smoothly interpolate between situations of different symmetry
as well as between different interactions. For example, $s_1 = 1, s_2 = \dots = s_k = 0$ (i.e., 
$N_1 = N$) corresponds 
to the extremal case of full symmetry, where only those terms in the interaction $V \in \V^\inv_k$ 
contribute that contain only projectors $P_1$ associated with the $N_1$-factor in the 
symmetry group. More generally, if $\Delta_l$ is a face of $\Delta_k$, 
$l<k$ then we have 
\ben
\label{129}
\sigma_{V, k} \restriction \Delta_l = \sigma_{V, l}, 
\een 
where it is understood that the ``$V$'' appearing in $\sigma_{V, l}$ is the 
formal expression in $\V^\inv_l$ obtained by dropping in $V \in \V^\inv_k$ all 
terms containing projectors that are not associated with the extremal points of $\Delta_l$. 
Using the restriction property~\eqref{129}, one can also construct bundles of interacting
field algebras over an arbitrary $k$-dimensional ($C^0$-) manifold $X$ by triangulating 
$X$ into simplices $\Delta_l$.

As in the case of full symmetry, the perturbative expansion of the interacting fields 
$\eps^{|\Phi|/2+T}\Phi_V$ defined via an interaction $V \in \V_k^\inv$ 
as an element of $\B_{V, \s}$, $\s = (s_1, \dots, s_k)$, is organized
in terms of Feynman graphs that are associated with Riemann surfaces.  Moreover, the
faces of these Feynman graphs defined by the closed index loops are now ``colored'' by 
the numbers $s_\alpha$. To illustrate this in an example, consider the 
interacting field $\eps^{3/2} (\tr \phi)_{\theta V}$ in 
case of 3 colors, $k = 3$, with interaction $\theta(x) V$, where
$\theta$ is a cutoff function, and where we take $V$ to be
\ben
\label{130}
V(\phi) = g \sum_{\alpha_i \in \{1, 2, 3\}}
\tr (P_{\alpha_1} \phi P_{\alpha_2} \phi \cdots P_{\alpha_n} \phi), 
\een
In order to make things a little more interesting, we restrict the sum in this equation to 
sequences of colors $(\alpha_1, \dots, \alpha_n)$ such that 
\ben
\label{131}
\alpha_1 \neq \alpha_2 \dots \neq \alpha_n \neq \alpha_1. 
\een
In the graphical notation introduced above this condition means that 
the vertices occurring in $V$ are restricted by the property that 
adjacent projectors $P_{\alpha_i}$ (as one moves around the vertex) are different. 

A retarded product appearing in the perturbative expansion of 
$\eps^{3/2}(\tr\phi)_{\theta V}$ with $V$ given by eq.~\eqref{130}, can now be written as a sum of 
contributions from individual Feynman graphs as follows:
\begin{multline}
\label{132}
\eps^{3/2} R(V(y_1) \cdots V(y_n); \tr \phi(x)) = 
g_{\rm t}^n \sum_{{\rm genera}\,\,H} \eps^H \sum_{{\rm graphs} \,\, \Gamma} \sum_{{\rm colorings} \,\, \C}
C_{\C, \Gamma} \cdot s_1^{F_1} s_2^{F_2} s_3^{F_3} \cdot \\
\cdot \eps^{f/2 + T} r_{\Gamma}(y_1, \dots, y_n; x) 
\,\,: \tr \prod_i \prod_{\alpha_i}\phi(y_i)P_{\alpha_i}               
\cdots \tr \prod_j \prod_{\beta_j}\phi(y_j)P_{\beta_j} :_H. 
\end{multline}
The notation used in the expression is analogous to that in the corresponding equation~\eqref{98}
in the case of full symmetry, with the following differences: The $r_\Gamma$ are 
the distributions appearing in the expansion of the interacting field in scalar 
$\phi^n$-theory. By contrast to eq.~\eqref{98},  
there appears now an additional sum over colorings, $\C$, over all 
ways to assign colors $s_1, s_2, s_3$ 
to those little surfaces in the big surface $\S$ associated with the Feynman graph 
not containing currents, in such a way that adjacent surfaces are never 
occupied by the same color (this corresponds to the property~\eqref{131} of $V$), 
and $F_\alpha$ is the number of such little surfaces colored by $s_\alpha$. The combinatorical 
factor $C_{\Gamma, \C}$ counts the number of ways in which a given coloring scheme can 
be produced by assigning the different terms in $V$ to the vertices. 
The external currents are again collected in the normal ordered
term appearing in eq.~\eqref{132}, where each trace corresponds to following through the 
index line to which the currents within that trace belong, but there now appear 
also the projectors $P_\alpha$ that are encountered when following through such an 
index line. A similar expansion can be written down for the interacting fields without 
cutoff $\theta$, as defined in eq.~\eqref{68}.

Thus, roughly speaking, the Feynman expansion of on interacting field $\eps^{3/2}(\tr \phi)_V$
with interaction $V$ given by~\eqref{130} 
differs from the corresponding expansion of $\phi_V$ in the theory of a single scalar field with 
$V = g\phi^n$ only in that the little surfaces in the graphs defined by the propagator lines 
are now colored according to the structure of the interaction $V$, 
and each coloring is weighted by the number $\prod_\alpha s_\alpha^{F_\alpha}$
where $F_\alpha$ is the number of little surfaces collored by $\alpha \in \{1, 2, 3\}$. 
The property~\eqref{131} of the interaction chosen in our example implies that only 
those graphs occur which can be colored by 3 colors in such a way that adjacent little
surfaces have different colors. In other words, there cannot appear any Feynman graphs such 
that the associated surface cannot be colored by less than 4 colors in this way. 
Thus, by choosing the interaction $V$ in the way described above, we have, in effect,
suppressed certain Feynman graphs that would be present in scalar $\phi^n$-theory.

\section{Summary and comparison to other approaches}

In this paper, we have constructed perturbatively the gauge invariant interacting quantum field operators 
for scalar field theory in the adjoint representation of $U(N)$, with an arbitrary 
gauge invariant interaction. These operators are members of an abstract algebra, whose structure
constants, as we demonstrated, have a well-defined limit as $N \to \infty$, or, more properly, 
are power series in $1/N$, with positive powers (provided the coupling parameters are also 
rescaled in a specific way by suitable powers of $1/N$). In this sense, these algebras, and the interacting quantum fields 
that are the elements of this algebra, also possess a well-defined large $N$ limit. We showed that
the renormalization group flow can be defined on the algebraic level via a 1-parameter family 
of isomorphisms acting on the fields via a rescaling of the spacetime arguments, the field strength, 
and an appropriate change in the coupling parameters. That flow was shown to be a power series
in $1/N$ with positive powers, and hence has a large $N$ limit. We also presented similar results in the case 
when the interaction of the fields is not invariant under $U(N)$, but only invariant under 
certain diagonal subgroups. We did not address issues related to the convergence of the perturbation
expansion or the expansion in $1/N$.

Our motivation for investigating the formulation of the $1/N$-expansion in an algebraic framework rather than 
via Green's functions of the vacuum state --- as is conventionally done --- was that the algebraic formulation 
is completely local in nature and thereby bypasses potential infra-red problems, which can
occur in the usual formulations via Green's functions in massless theories. Also, although we explicitly
only worked in Minkowski space, we were strongly motivated by the fact that an algebraic approach is essential 
if one wants to formulate quantum field theory in a generic curved spacetime, where no preferred vacuum
state exists. Actually, since our arguments are mostly of combinatorical nature, 
we expect that the present algebraic formulation of the $1/N$ expansion can
be carried over rather straightforwardly to curved space.

The algebraic approach presented in this paper is rather different in appearance from the usual formulation
via Green's functions, so we would briefly like to explain the relationship between the two approaches.
In the conventional approach, one considers the $N$-dependence of the vacuum Green's functions\footnote{One 
normally considers time ordered Green's functions, but the arguments do not depend on the time ordering
and therefore also equally apply to the Wightman functions, which we prefer to consider here.} of gauge
invariant interacting fields, $G_n = \omega_0(\Phi_V \dots \Phi_V)$ associated with the interaction $V$. For
example, for $\Phi = \tr \phi^2$, one finds that the corresponding connected Green's function $G_n^{\rm conn}$
receives contributions of order $N^{2-2H}$ from Feynman graphs of genus $H$ (assuming that the couplings
in $V$ are scaled by appropriate powers of $1/N$). Thus, the planar diagrams $H=0$
make the leading contribution at large $N$, with $G_n^{\rm conn} \sim N^2$, independent of $n$.

If one wants to reconstruct from the Green's functions 
the Hilbert space of the theory and the interacting field observables as linear operators
on that Hilbert space, one needs to consider not the connected Green's functions, but the 
Wightman Green's functions $G_n$ themselves, since the latter enter in the Wightman reconstruction argument. 
Writing the Wightman Green's functions $G_n$ in terms of $G_n^{\rm conn}$ via the usual formulae, one
immediately gets that $G_n \sim N^{2n}$. Hence, it is clear that, if 
one wants these constructions to be well defined at infinite $N$, then one needs to consider the 
normalized fields $N^{-2} \tr \phi^2$. Similar remarks also apply to more general composite fields,
with appropriate powers of $1/N$ in the normalization factor, 
depending on the number of traces and the number of basic fields. These powers coincide precisely with 
the powers found in our algebraic approach (see eq.~\eqref{93}) by different means. The arguments that we have just 
given are of course only formal, because the reconstruction theorem, as it stands,  
is not really applicable in perturbation theory. Also, as we have already emphasized several times, 
the $G_n$ may actually be ill defined because they may involve infra-red divergent integrations over
interaction vertices (in massless theories). The methods of this paper, on the other hand, give a
rigorous construction of the field theory at the algebraic level that, by contrast to the formulation
via Green's functions,  
should also be applicable in curved spacetimes. 

It is a trivial consequence of the large $N$-behavior of the connected Green's functions that, 
in the large $N$ limit, the Wightman Green's functions $\hat G_n$ of the suitably normalized field 
operators factorize into 1-point functions, $\hat G_n(x_1, \dots, x_n) \sim \hat G_1(x_1) \cdots \hat G_1(x_n)$.
Thus, it is formally clear that the large $N$ theory is {\em abelian}, i.e., the field commutators 
vanish. This can be seen explicitly in our algebraic framework, since the commutator of any two 
(suitably normalized) interacting fields is seen to be of order $N^{-2}$. Thus, the algebra $\A_V$ of 
interacting fields is abelian in the large $N$ limit, and consequently the representations are degenerate. 
This behavior can be nicely formalized in the algebraic framework by viewing $\A_V$ as a {\em Poisson algebra}\footnote{
A Poisson algebra is an algebra $\A$ together with an antisymmetric bracket 
$\{ \, . \, , \, . \, \}$ from $\A \times \A$ to $\A$ satisfying the Leibniz rule
$\{ab, c\} = a\{b, c\} + \{a, c\} b$, together with the Jacobi identity. A Poisson algebra
is called abelian if $\A$ is abelian. The observables of a classical field theory form 
an abelian Poisson algebra, with the (commutative) algebra multiplication given by 
pointwise multiplication of the observables, and with the Poisson bracket given in 
terms of the symplectic structure of the theory. A trivial example of a nonabelian
Poisson algebra is any non-commutative algebra with the Poisson bracket defined by the algebra commutator.
}, with antisymmetric bracket defined by $\{ \, . \, , \, . \, \}
= N^2 [ \, . \, , \, . \, ]$. In the limit of large $N$, that Poisson algebra becomes abelian, as is 
also the case for the classical limit\footnote{The classical limit and the large $N$ limit are not, 
of course, equivalent.} $\hbar \to 0$ (the appropriate definition of the Poisson bracket in that 
case being $\{ \, . \, , \, . \, \} = (i\hbar)^{-1} [ \, . \, , \, . \, ]$). Thus, 
it is seen clearly at the algebraic level that there exist formal similarities between the 
large $N$ limit and the classical limit, and that, in particular, the large $N$ limit 
of a field theory does not define a quantum field theory in the usual sense, but rather
a Poisson algebra. We note that the large $N$ limit can thereby be interpreted, within our 
algebraic framework, as some kind of ``deformation quantization''~\cite{deform}, 
the deformation parameter being $1/N^2$.

\vspace{2cm}
{\bf Acknowledgements:} I would like to thank K.~H.~Rehren for useful conversations. 
This work was supported by NSF-grant~PH00-90138 to the University of Chicago.

\end{document}